\documentclass[]{amsart}

\usepackage{epsfig}
\usepackage{graphics}
\usepackage{graphicx}
\usepackage{amsmath}
\usepackage{url}
\usepackage{subfigure}

\pdfoutput=1

\begin{document}

\title[Voronoi diagram on Voronoi automata]{Vesicle computers:\\ Approximating Voronoi diagram\\ on Voronoi automata}

\author[Adamatzky]{Andrew Adamatzky}
\address[A.~Adamatzky]{Unconventional Computing Centre, University of the West of England, Bristol, United Kingdom}

\author[De Lacy Costello]{Ben De Lacy Costello}
\address[B.~De Lacy Costello]{Centre for Analytical Chemistry and Smart Materials,sity of the West of England, Bristol, United Kingdom}

\author[Holley]{Julian Holley}
\address[J.~Holley]{Unconventional Computing Centre, University of the West of England, Bristol, United Kingdom}

\author[Gorecki]{Jerzy Gorecki}
\address[J.~Gorecki]{Institute of Physical Chemistry PAN, Warsaw, Poland}

\author[Bull]{Larry Bull}
\address[L.~Bull]{Artificial Intelligence Group, University of the West of England, Bristol, United Kingdom}

\date{\today}

%\pacs{82.40.-g; 82.40.Ck; 89.75.Kd; 89.75.Fb; 89.20.Ff}

\begin{abstract}
Irregular arrangements of vesicles filled with excitable and precipitating chemical systems 
are imitated by Voronoi automata --- finite-state machines defined on  a planar Voronoi diagram.  
Every Voronoi cell takes four states: resting, excited, refractory and precipitate. A resting 
cell excites if it has at least one excited neighbour; the cell  precipitates if a ratio of 
excited cells in its neighbourhood to its number of neighbours exceed certain threshold. To 
approximate a Voronoi diagram on Voronoi automata we project a planar set onto automaton lattice, 
thus cells corresponding to data-points are excited. Excitation waves propagate across the Voronoi automaton, 
interact with each other and form precipitate in result of the interaction. Configuration of precipitate 
represents edges of approximated Voronoi diagram. We discover relation between quality of Voronoi diagram 
approximation  and precipitation threshold, and demonstrate  feasibility of our model in approximation Voronoi diagram of arbitrary-shaped objects and a skeleton of a planar shape. 

\vspace{0.5cm}

\noindent
\emph{Keywords: Voronoi diagram, automata, discrete networks, wave dynamics} 
\end{abstract}

\maketitle

\section{Introduction}

A regular, or irregular but manually designed, arrangements of vesicles filled with excitable chemical
mixtures bear huge computational potential~\cite{adamatzky_BZBALLS_2010}. When vesicles are in close, at least in a diffusion terms, contact with each other, via tiny pores, excitation waves can pass from one vesicle to its close neighbour. Excitation wave-fragments keep their shape, more or less constant, inside each BZ-vesicle. A wave-fragment passing from one
vesicle to another it contracts, due to the restricted size of the connecting pore. When two or more wave-fragments collide inside
a vesicle they can annihilate, deviate, or multiply. When interpreting presence/absence of wave-fragments in any given as a value of Boolean variable we can implement all basic operations of a Boolean logic via collisions between wave-fragments in a vesicle. 
In computer experiments we designed a binary adder in a hexagonal array of vesicles filled with excitable chemical mixture~\cite{adamatzky_BZBALLS_2010}, built polymorphic logical gates (switching between {\sc xnor} and {\sc nor}) by using illumination to control outcomes of inter-fragment collisions~\cite{adamatzky_polymorphic_2010} and geometry-modulated complex arithmetical circuits~\cite{holley_2010}. Our theoretical ideas and results got experimental chemical laboratory back up -- results on information transfer between Belousov-Zhabotinsky mixture enclosed in lipid membrane are successful~\cite{gorecki_private,neuneu}.

\begin{figure}[!tbp]
\centering
\subfigure[]{\includegraphics[width=0.59\textwidth]{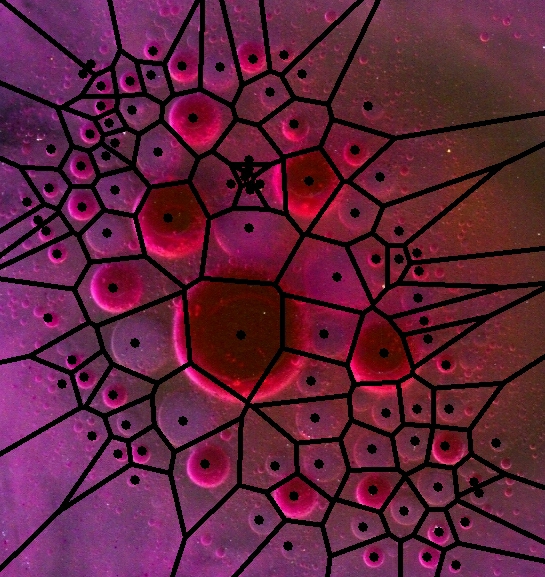}}
\subfigure[]{\includegraphics[width=0.39\textwidth]{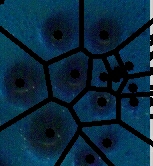}}
\subfigure[]{\includegraphics[width=0.59\textwidth]{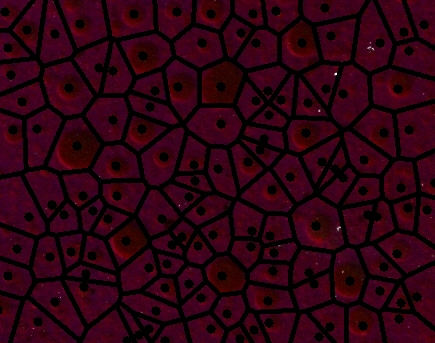}}
\caption{Examples of Voronoi representation of bubble conglomerates: 
(a)~rhodamin, vegetable oil, water and silicon oil (polydimethylsiloxane), 
(b)~blue food coloring and vegetable oil, 
(c)~rhodamin, water and silicon oil mixtures.}
\label{examples}
\end{figure}

While in computer models regular arrangement of uniform vesicles is effortless real-life experiments bring nasty surprises. Usually vesicles are different sizes, they do not form a hexagonal lattice as a rule, and they may be unstable, coalescence transforms 
fine-grained networks of elementary vesicle-processors into a coarse-grained assembles of monstrous vesicular structures. What kind of computation can be done on an irregular arrangement of non-uniform vesicles? We address the question by representing the vesicle assembles by automata networks and studying how planar subdivision, Voronoi diagram, problem can be solved in such vesicle-automata. 

We abstract vesicle assembles as planar Voronoi diagrams (Fig.~\ref{examples}) of planar sets, points of which are centres of the vesicles. Voronoi diagram is routinely as approximation of arrangements of discs~\cite{gervois_1995} and sphere packing~\cite{lochman_2006,filatovs_1998,luchnikov_2002}. The diagram is also used in structural analysis of liquids and gases~\cite{anikeenko_2004}, and protein structure~\cite{poupon_2004}, and to model dense gels~\cite{zarzycki_1992} and inter-atomic bonds~\cite{hobbs_1995}. The Voronoi diagrams are introduced in Sect.~\ref{voronoidiagram}. We assume that every cell of a Voronoi diagram is a finite-state that takes four states and updates its states depending on states of its first and second order neighrbours. We design a cell-state transition function which combines generalised, and highly-abstracted, properties of both excitable and precipitating chemical media: a local disturbance gigves birth to quasi-circular waves of excitation while collisions between the waves lead to precipitation. The Voronoi automaton is defined in Sect.~\ref{voronoiautomata}. 

The problem solved by Voronoi automats is the approximation of Voronoi diagram. Approximated Voronoi diagram is much more coarse-grained than Voronoi diagram on which excitable-precipiating automaton is built. To approximate a Voronoi diagram on Voronoi automata we project a planar set onto automaton lattice, thus cells corresponding to data-points are excited. Excitation waves propagate across the Voronoi automaton, interact with each other and form precipitate in result of the interaction. Configuration of precipitate represents edges of approximated Voronoi diagram. In our model precipitation depends on a local density of excitation --- precipitation threshold. For low precipitation threshold the medium becomes cluttered with meaningless clusters of precipitate, for high threshold few domain of precipitation is formed. Our quest for optimal threshold of precipiation is narrated in Sect.~\ref{approximation}. In Sect.~\ref{shapes} we show optimal threshold found can be equally used to approximate Voronoi diagram of arbitrary planar shapes and also skeleton of planar shape.

\section{Voronoi diagram}
\label{voronoidiagram}

Let $\bf P$ be a nonempty finite set of planar points. A planar Voronoi diagram~\cite{voronoi_1907} of
the set $\bf P$ is a partition of the plane into such regions, that for any element
of $\bf P$, a region corresponding to a unique point $p$ contains all those points of
the plane which are closer to $p$ than to any other node of $\bf P$. A unique region
$vor(p) = \{z \in {\bf R}^2: d(p,z) < d(p,m) \forall m \in {\bf R}^2, \, m \ne z \}$
assigned to point $p$ is called a Voronoi cell of the point $p$~\cite{preparata_shamos_1985}. The boundary 
$\partial vor(p)$  of
the Voronoi cell of a point $p$ is built of segments of bisectors separating
pairs of geographically closest points of the given planar set $\bf P$. A union of
$VD({\bf P}) = \cup _{p \in {\bf P}} \partial vor(p)$
all boundaries of the Voronoi cells determines the planar Voronoi diagram~\cite{preparata_shamos_1985}.
A variety of Voronoi diagrams and algorithms of their construction can be found in \cite{klein_1990, okabe_2000}.

Approximation of Voronoi diagrams with propagating patterns is based on time-to-distance transformation:
to approximation a bisector separating planar points $p$ and $q$ we initiate growing patterns at $p$ and $q$. 
The pattern travel the same distance from the sites of origination before they meet each other, The loci 
where the waves meet indicate sites of the computed bisector~\cite{adamatzky_1994,adamatzky_1996}. Precipitating reaction-diffusion chemical media are proved to be an ideal computing substrate for approximation of the planar Voronoi diagram~\cite{tolmachev_1996,ben_2004,adamatzky_rdc}. A Voronoi diagram can be approximated in a two-reagent medium. 
One reagent $\alpha$  is saturated in the substrate, drops of another reagent $\beta$ are applied to the sites corresponding to planar points to be separated by bisectors. The reagent $\beta$ diffuses in the substrate and reacts with reagent $\alpha$. Colored precipitate is produced in the reaction between $\alpha$ and $\beta$. When two or more waves of diffusing $\alpha$ meet, no precipitate is formed~\cite{ben_2004}. Thus uncolored loci of the reaction-diffusion medium represent bisectors of the 
computed diagram.  A range of chemical precipitating processors is designed and working prototypes are tested in laboratory 
conditions ~\cite{tolmachev_1996,ben_2003a,ben_2003b,ben_2004,ben_2009}.

\section{Voronoi automata}
\label{voronoiautomata}

A Voronoi automaton is  a tuple $\mathcal{V}=\langle {\mathbf V}({\mathbf P}), {\mathbf Q}, {\mathbf N}, u, f  \rangle$,
where $\mathbf P$ is a finite planar set, ${\mathbf V}({\mathbf P}) = \{V(p): p \in {\mathbf P}  \}$,
$\mathbf Q$ is finite set, $\mathbf N$ is a set of natural numbers and 
$u: {\mathbf V}({\mathbf P}) \rightarrow {\mathbf V}({\mathbf P})^k$ is second-order neighbourhood, 
$0 < k < |{\mathbf P}|$, and $f: {\mathbf Q}^k \rightarrow {\mathbf Q}$ is a cell-state transition function.

Excitable-precipitating Voronoi automata studied in present paper are specified as follows. 
Cell state set has for elements, ${\mathbf Q}=\{ \circ, +, -, \# \}$. Thus we assign 
three excitation-related states --- resting ($\circ$), excited ($+$) and refractory ($-$) --- 
to cells, and one precipitate state $\#$. Cells update their state in discrete time. A state of cell $V(x)$ are time step 
$t \in \mathbf N$ is denoted as $V(x)^t$. All cells update their states in parallel using the same cell-state 
transition function. 

\begin{figure}[!tbp]
\centering
\subfigure[]{\includegraphics[width=0.7\textwidth]{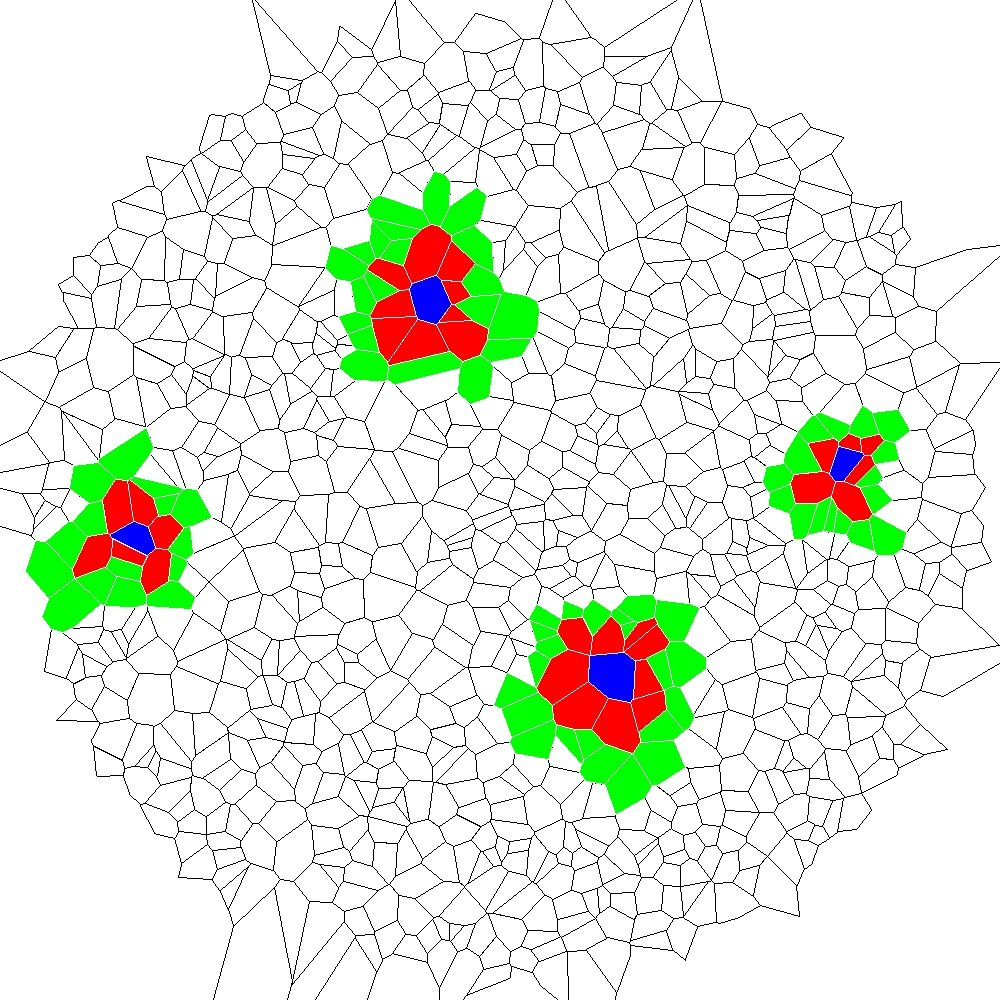}}
\subfigure[]{\includegraphics[width=0.7\textwidth]{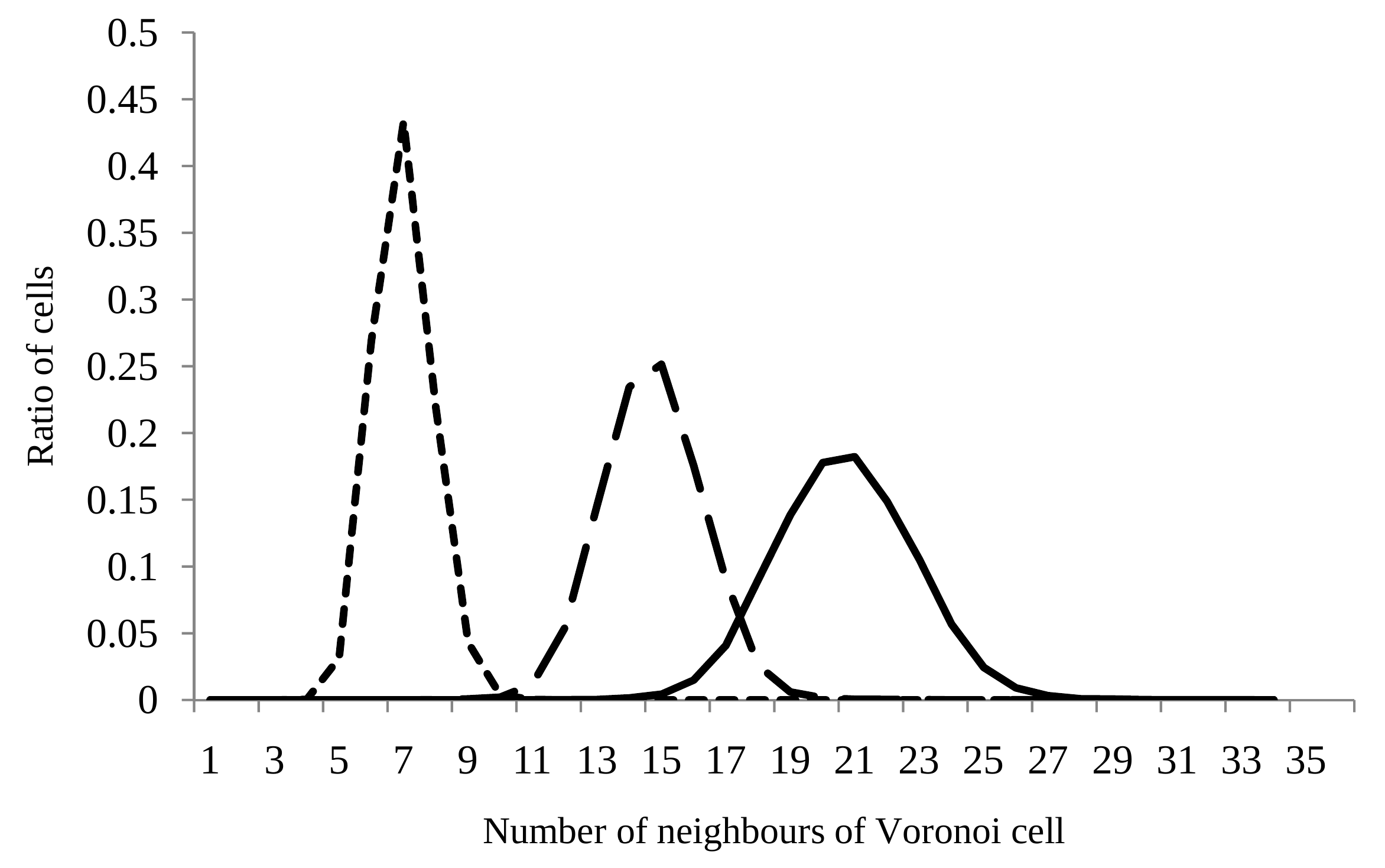}}
\caption{Structure and properties of neighbourhoods. 
(a)~Example of Voronoi tesselation with second-order neighbourhood structures highlighted for four cells. 
In each case central cell $p$ is filled with blue (black in gray-scale reproduction) colour, first-order neighbours of $w(p)$ 
are filled with red (dark-gray) colour, and neighbours of $w(p)$ are filled with green (light-gray) colour. For each blue (black)
Voronoi cell a second-order neighbourhood is a set of red (dark-gray) and green (light gray) Voronoi cells.
(b)~Distribution of first order, or immediate, neighbours (dotted line), second order neighbours (dashed line) and size of 
second order neighbourhood (solid line) in Voronoi automata of 15K cells.
} 
\label{neighbourhood}
\end{figure}

Let $w(p)$, $p \in {\mathbf P}$, be a first-order neighbourhood of a Voronoi cell $Vor(p)$ from ${\mathbf V}({\mathbf P})$:
$w(p) = \{V(q) \in {\mathbf V}({\mathbf P}):  \partial V(p) \cap \partial V(q) \}$, i.e. a set of Voronoi cells which 
have common edges with $V(x)$. A second-order neighbourhood $u(p)$ is a set of neighbours of first-order neighbours of 
$V(p)$: $u(p) = \bigcap_{V(q) \in w(V(p))} w(V(q)$. Examples of Voronoi cell neighbourhoods are shown 
in Fig.~\ref{neighbourhood}a. Distributions of numbers of immediate and second neighbours and sizes of second-order neighbourhood $u$ are given in Fig~\ref{neighbourhood}b. Predominant number, over 40\%, of cells have six first-order (immediate) 
neighbours. Almost half of the cells have 13 or 14 second-order neighbours. For any cell size of second-order neighbourhood $u$ is 
a sum of first-order neighbours and second-order neighbours. For Voronoi automata with high-density packing of 15K cells, 
half of the cells have either 19, or 20 or 21 neigbours. If we consider only immediate neighbourhood then $\mathcal V$ can be,
in principle, seen as a slightly distorted, hexagonal lattice. However,  by adopting second-order neighbourhood we 
are moving to a random structure with higher than hexagonal lattice coordinate number (a node in hexagonal lattice has 18-node second-order neighbourhood while dominating neighbourhood sizes in $\mathcal V$ are 19, 20 and 21).   

Transition form excited to refractory state is unconditional, i.e. takes place with regards to states of a cell's 
neighbours. A resting cell excites if it has at least one excited neighbour. 
In this particular model we take a refractory state and precipitate states are absorbing: one a cell takes either of these two state it does not update its state any longer. 

Neighbourhood sizes may differ between Voronoi cells therefore we use a state transition function, 
where a cell updates its state depending on a relative excitaton in its neghbourhood. We assume that 
precipitation occurs in a resting cell when a ratio of excited neighbours to a total number of 
neighbours exceeds some threshold $\eta \in [0,1]$. Let $\sigma(V(x)^t)$ be a number of excited cells in the 
cell $V(x)$'s second-order neighbourhood, $\sigma(V(x)^t) = \sum_{V(y) \in u(V(x))} |\{V(y): V(y)^t=+\}|$ then 
a cell updates its state by the following rule:   
$$
V(x)^{t+1}=
\begin{cases}
\#, \text{ if } V(x)^t=\circ \text{ and } \sigma^t(x)/\nu(x)>\eta\\
+, \text{ if } V(x)^t=\circ \text{ and }  \sigma^t(x)/\nu(x)>1\\
-, \text{ if } V(x)^t=+\\
\circ, \text{ otherwise }
\end{cases}
$$

In computational experiments we construct set $\mathbf P$ by fill a disc-container of radius 480 units 
with up to 15K points. The points are packed at random but there is at least 5 units distance between
any two points. Voronoi diagram $\mathcal V$ --- on which Voronoi automaton is constructed -- is calculated 
by a classical sweepline algorithm~\cite{fortuna_1986}.

\section{Constructing Voronoi diagram on Voronoi automata}
\label{approximation}

\begin{figure}[!tbp]
\centering
\subfigure[$t=1$]{\includegraphics[width=0.49\textwidth]{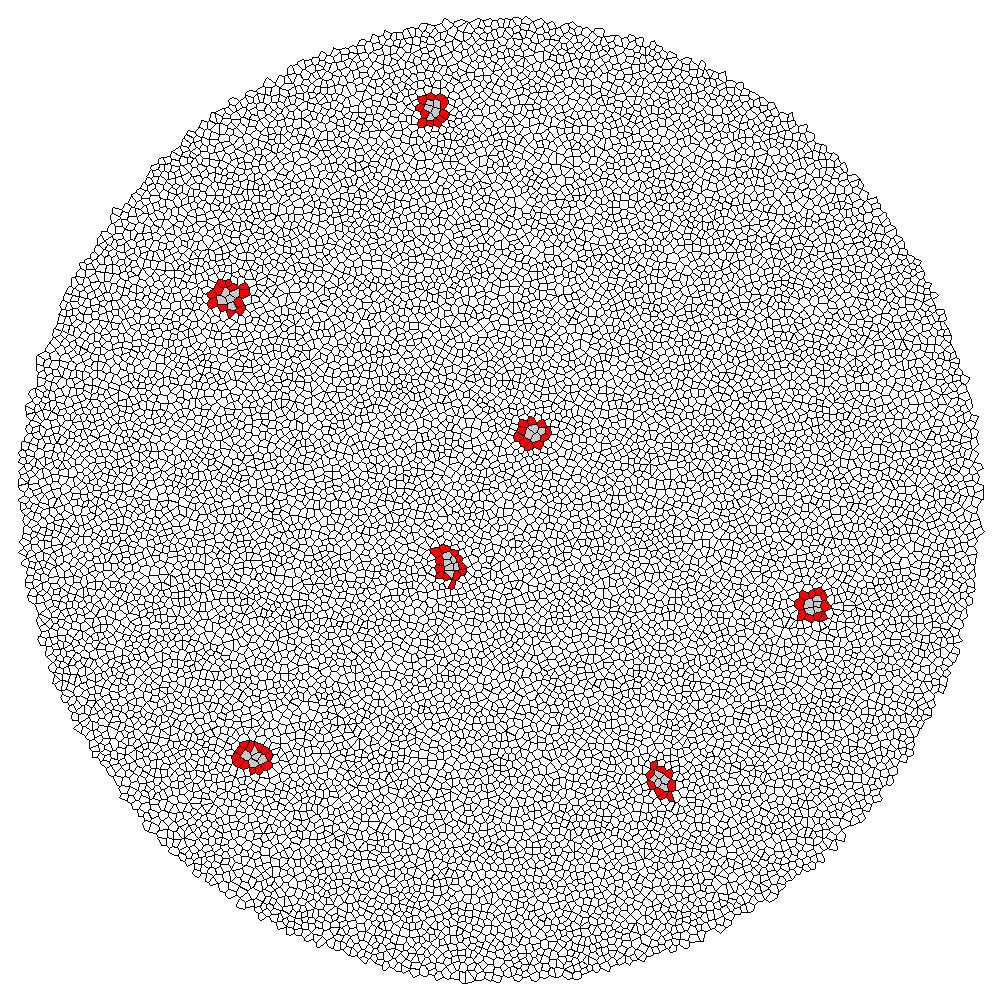}}
\subfigure[$t=12$]{\includegraphics[width=0.49\textwidth]{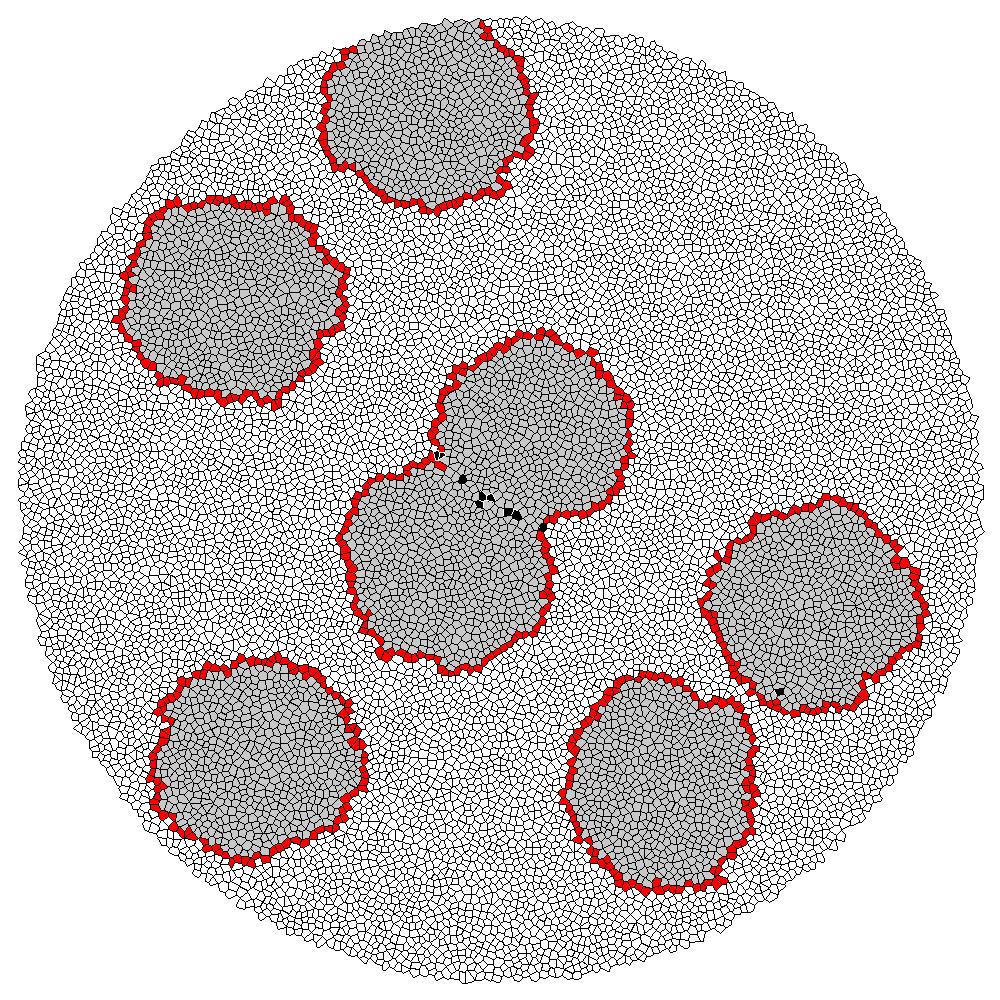}}
\subfigure[$t=20$]{\includegraphics[width=0.49\textwidth]{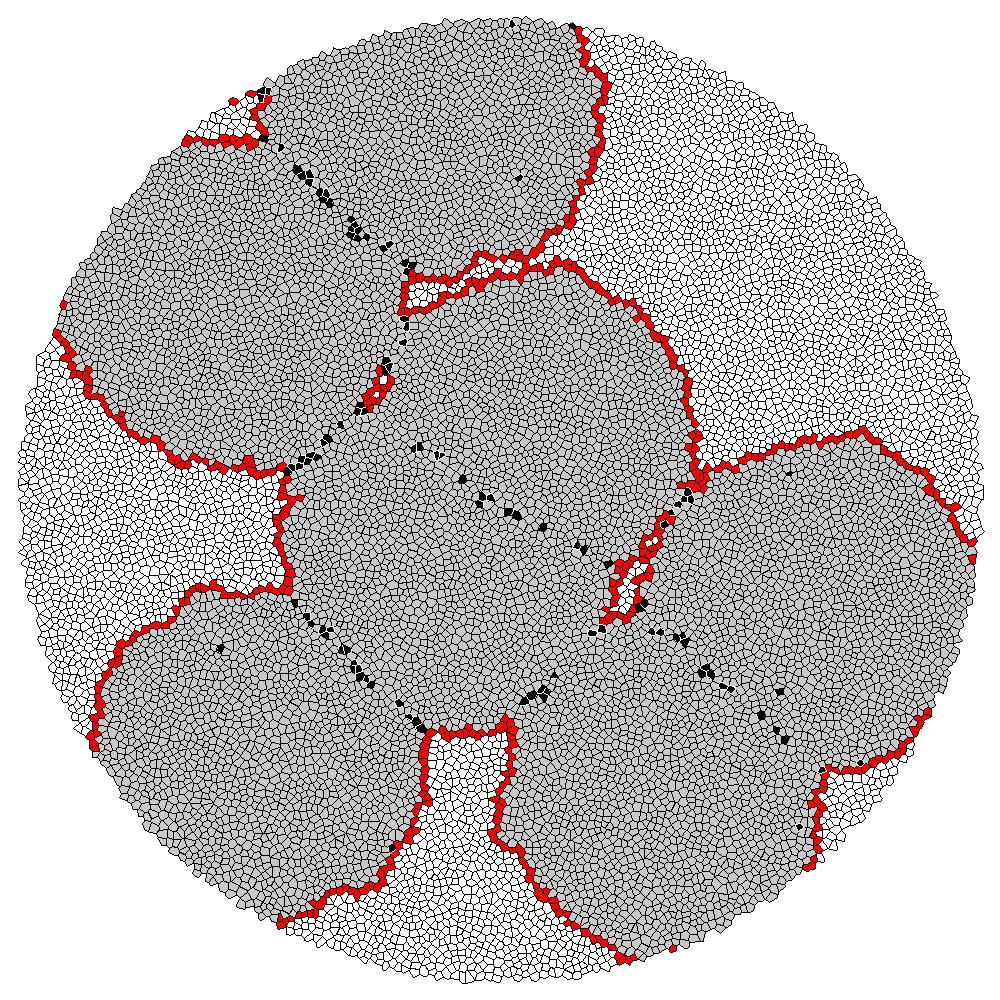}}
\subfigure[$t=44$]{\includegraphics[width=0.49\textwidth]{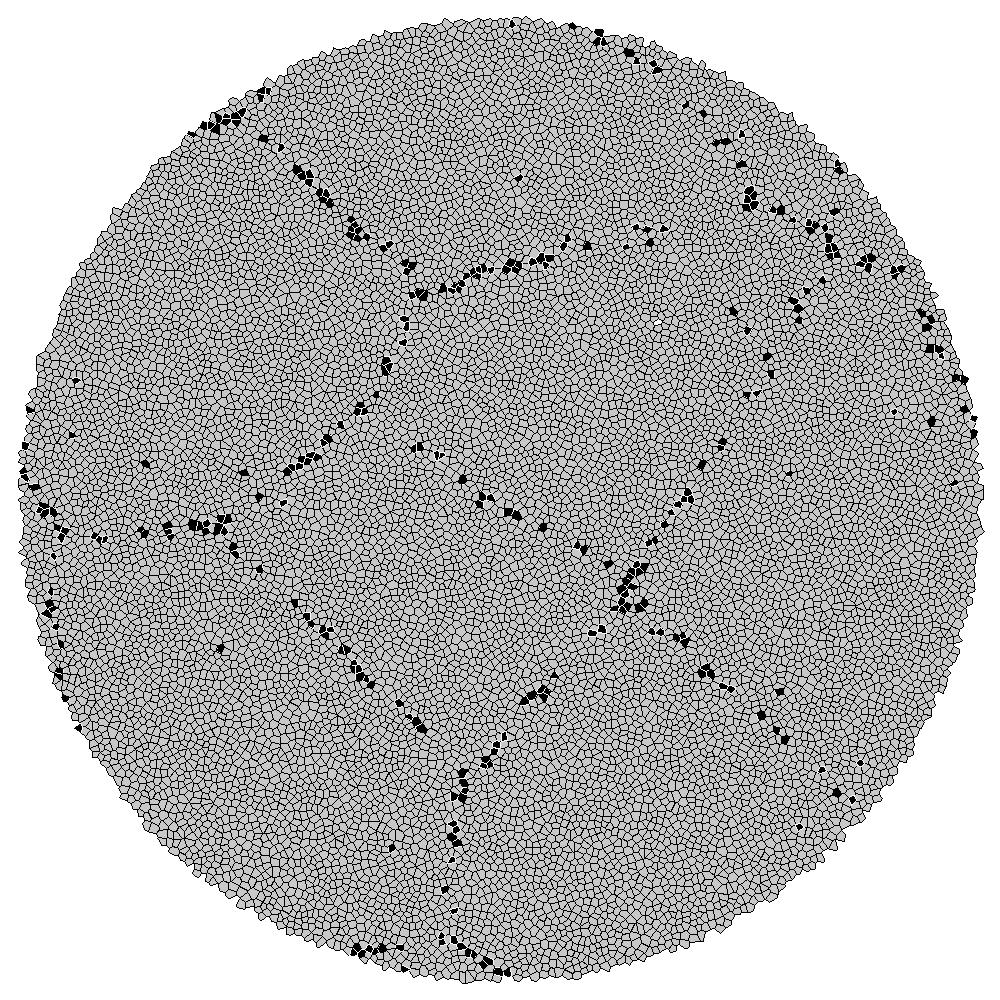}}
\caption{Approximation of Voronoi diagram $\mathfrak{V}(\mathbf{B})$ on Voronoi automaton $\mathcal V$, $\eta=0.4$. Points of 
$\mathbf B$ are projected onto $\mathcal V$ at time step $t=0$. Seven Voronoi cells of $\mathcal V$ are excited
and generated quasi-circular excitation waves. Configuration of $\mathcal V$ at time step $t=1$, when wave have just 
started to develop, is shown in (a). The waves propagate outwards their initial stimulation sites and covert Voronoi cells they are occupying into refractory states (bc). When two or more waves collide precipitation occurs. By 44th step of the automaton
development excitation extincts but domains of precipitate-cells represent edges of the approximated Voronoi diagram $\mathcal V$.
Resting cells are blank, excited cells red (dark gray), refractory cells gray, and precipitate cells black. } 
\label{points}
\end{figure}

Let $\mathbf B$ be a set of planar points on which a Voronoi diagram $\mathfrak{V}$ on $\mathcal V$, $\eta=0.4$, is approximated. We project $\mathbf B$ onto $\mathcal V$ and excite cells of $\mathcal V$ which are closer than 9 units to points of $\mathbf B$. Excitation waves spread  on $\mathcal V$. The excitation waves collide and precipitation occurs nearby sites of the waves' collisions. Configuration of cells in precipitate state represents edges of  $\mathfrak{V}(\mathbf{B}$. An example of excitation-precipitation dynamics in $\mathcal V$ which approximates $\mathfrak{V}(\mathbf{B})$, where $\mathbf B$ is a planar set of seven points, is shown in Fig.~\ref{points}. A scheme of the Voronoi diagram computed is shown in Fig.~\ref{scheme}a.

\begin{figure}[!tbp]
\subfigure[VD]{\includegraphics[width=0.24\textwidth]{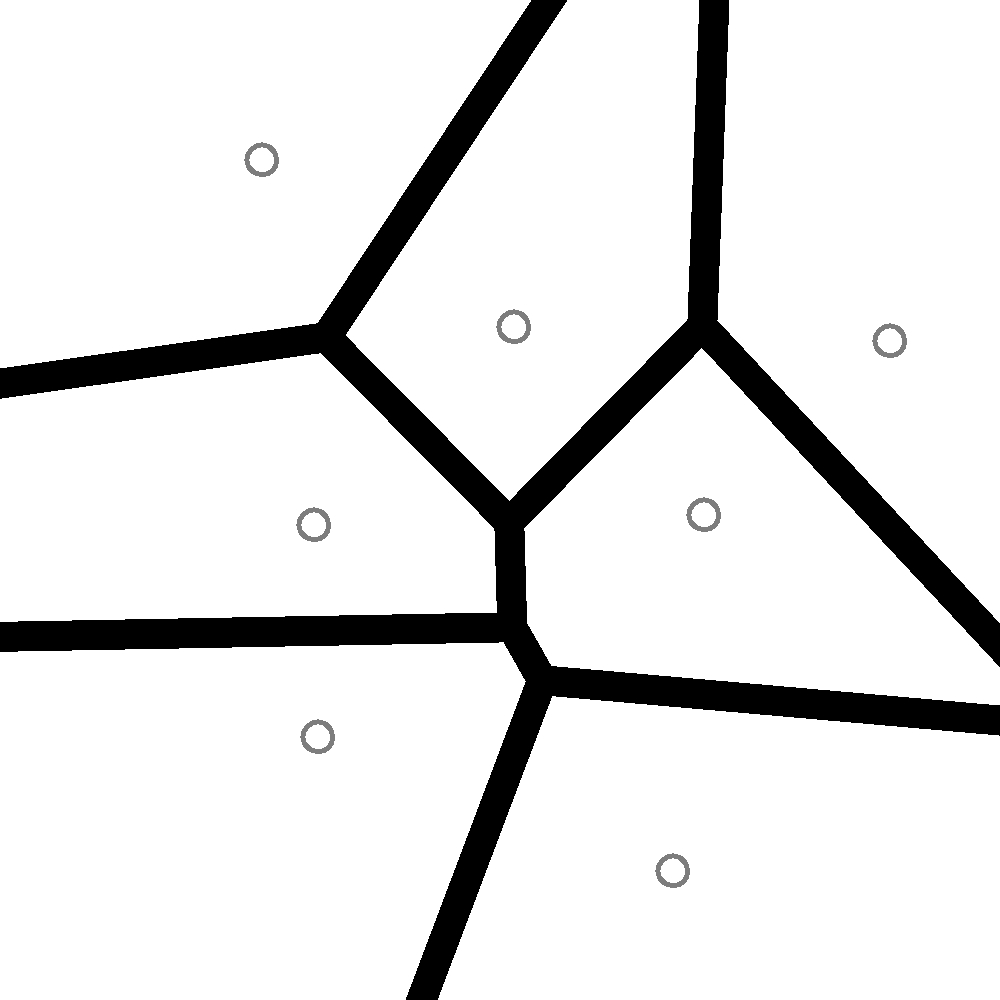}}
\subfigure[$\eta=0.2$]{\includegraphics[width=0.24\textwidth]{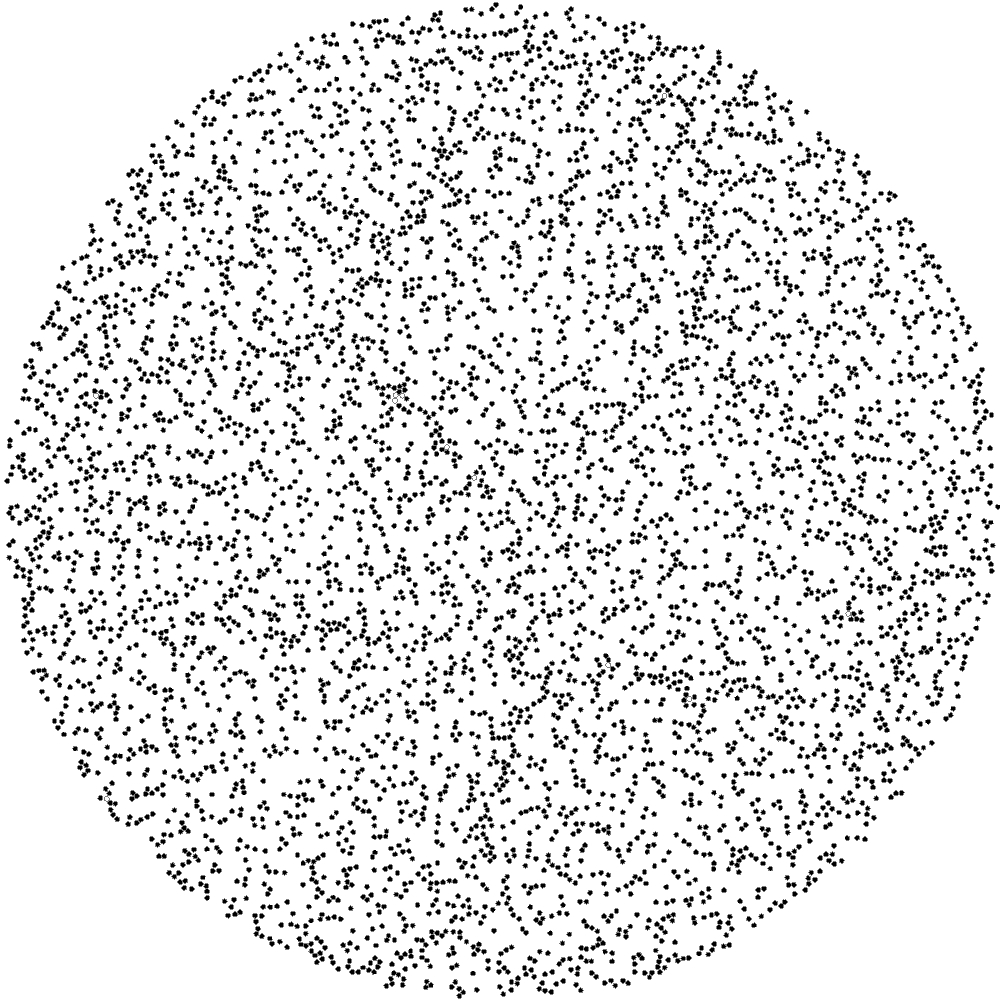}}
\subfigure[$\eta=0.225$]{\includegraphics[width=0.24\textwidth]{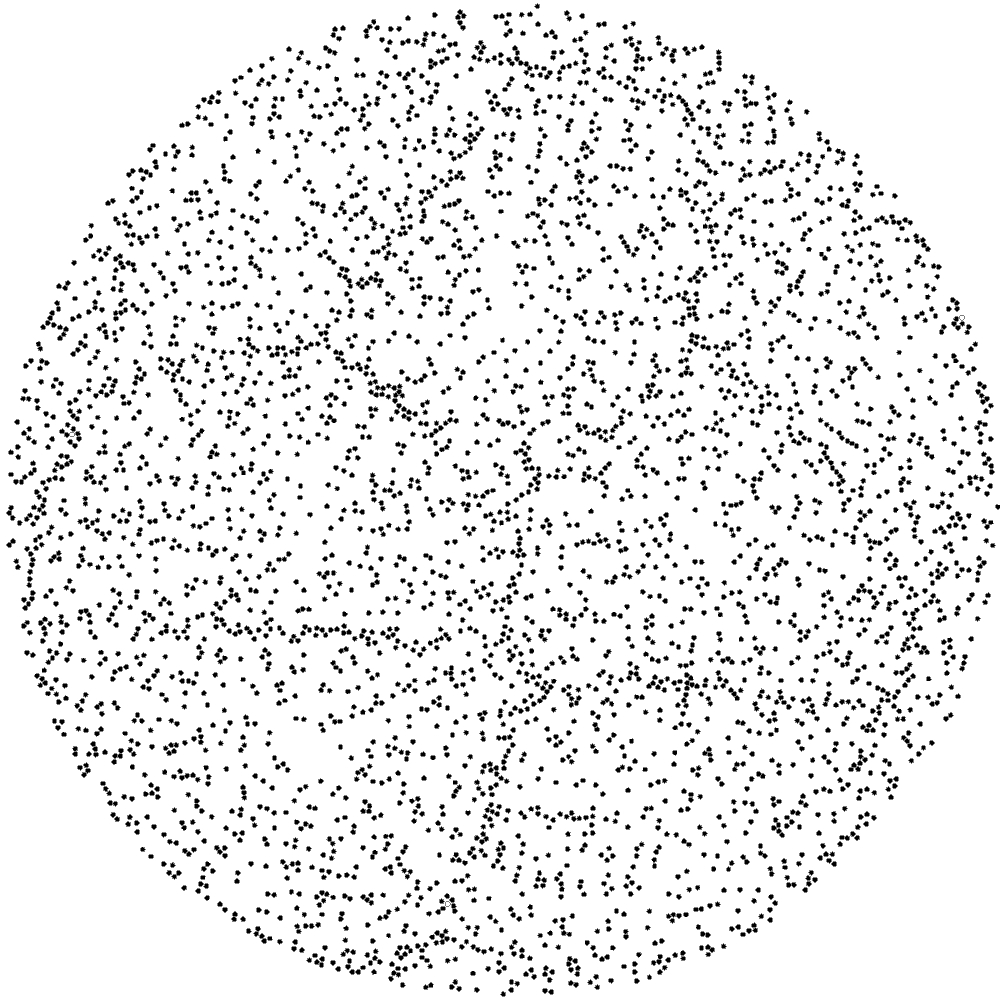}}
\subfigure[$\eta=0.25$]{\includegraphics[width=0.24\textwidth]{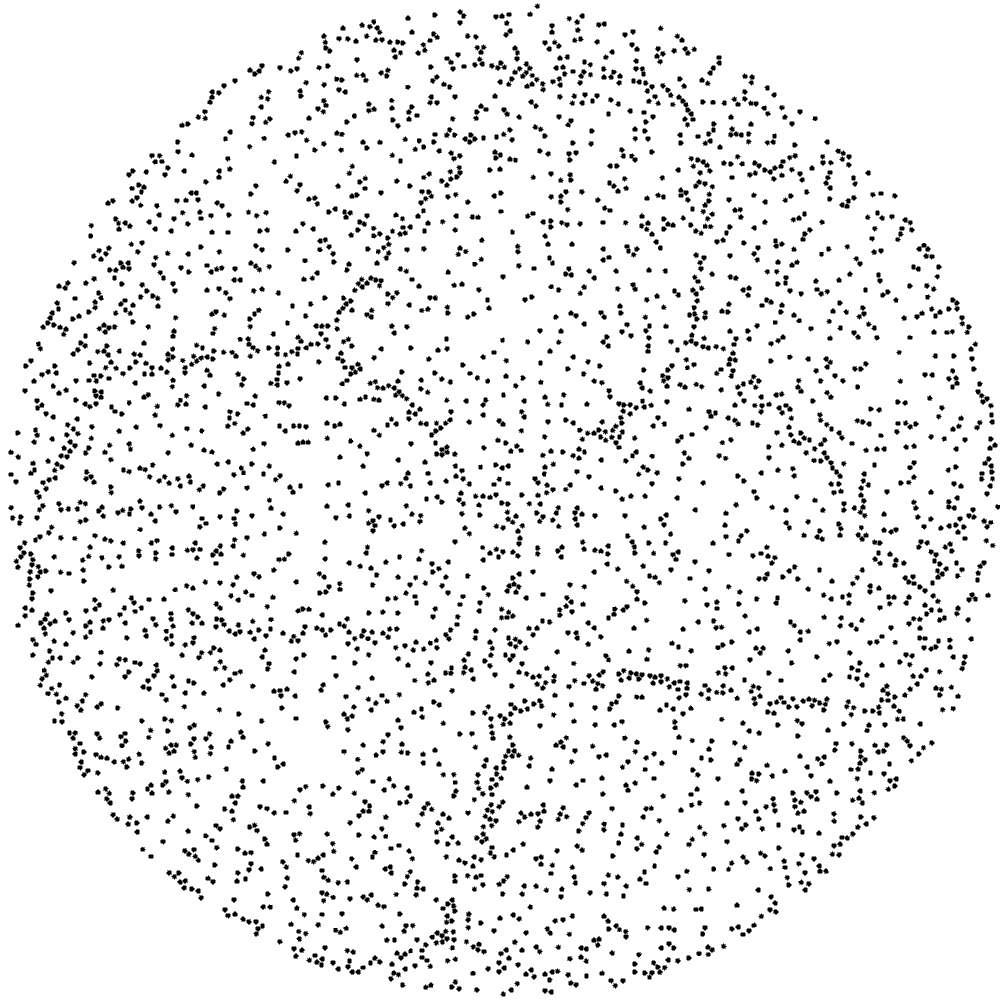}}
\subfigure[$\eta=0.275$]{\includegraphics[width=0.24\textwidth]{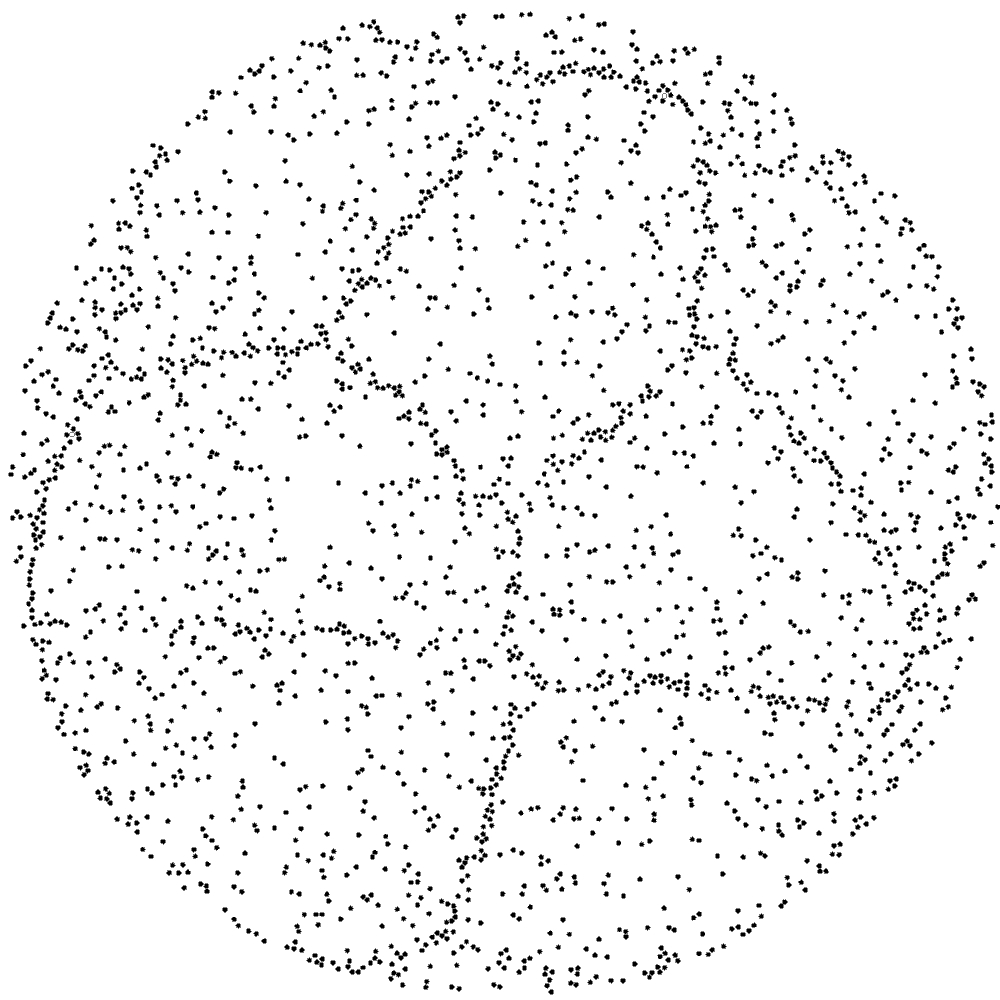}}
\subfigure[$\eta=0.3$]{\includegraphics[width=0.24\textwidth]{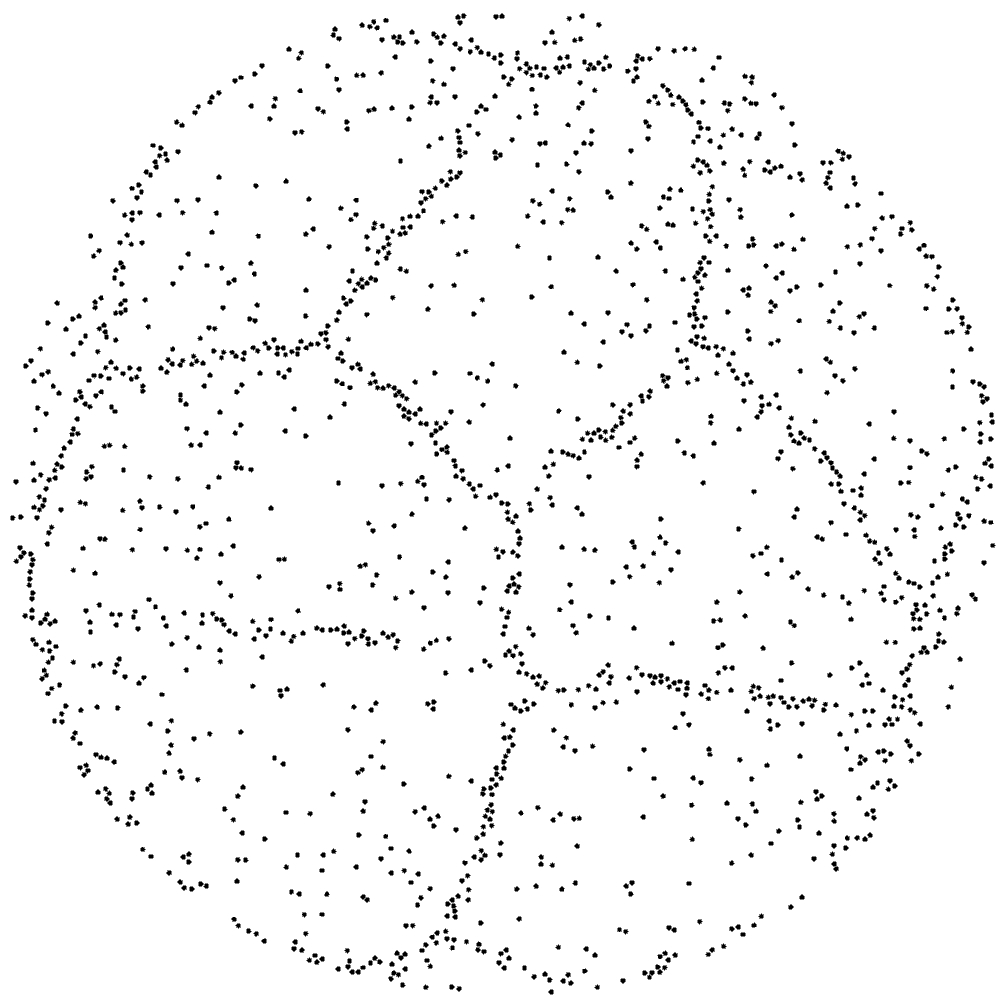}}
\subfigure[$\eta=0.325$]{\includegraphics[width=0.24\textwidth]{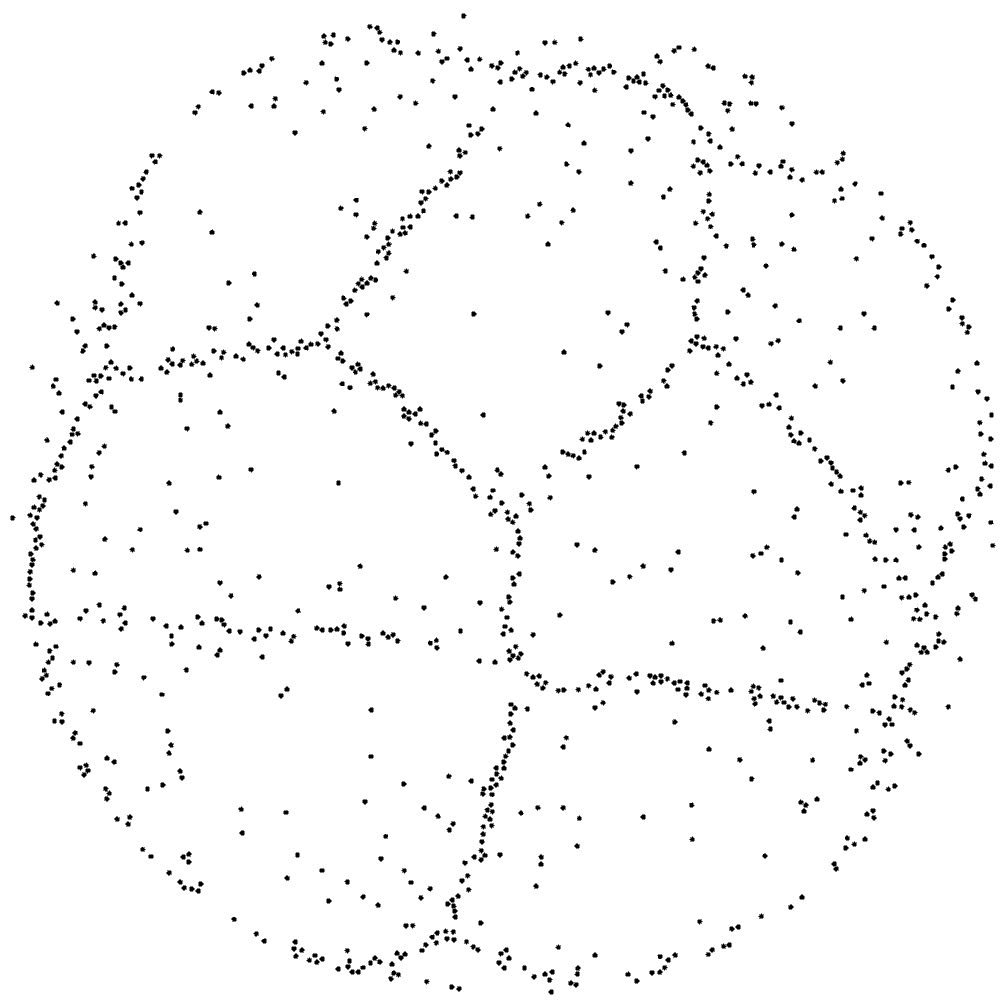}}
\subfigure[$\eta=0.35$]{\includegraphics[width=0.24\textwidth]{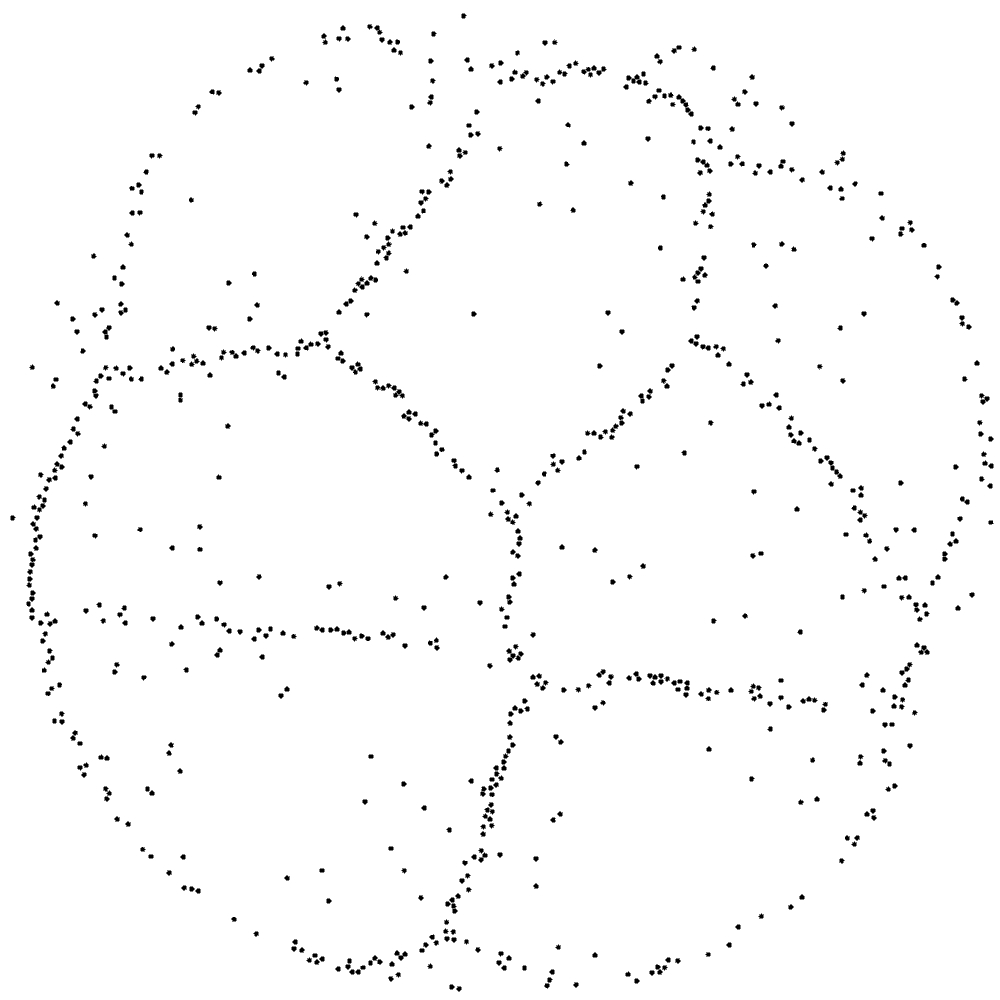}}
\subfigure[$\eta=0.375$]{\includegraphics[width=0.24\textwidth]{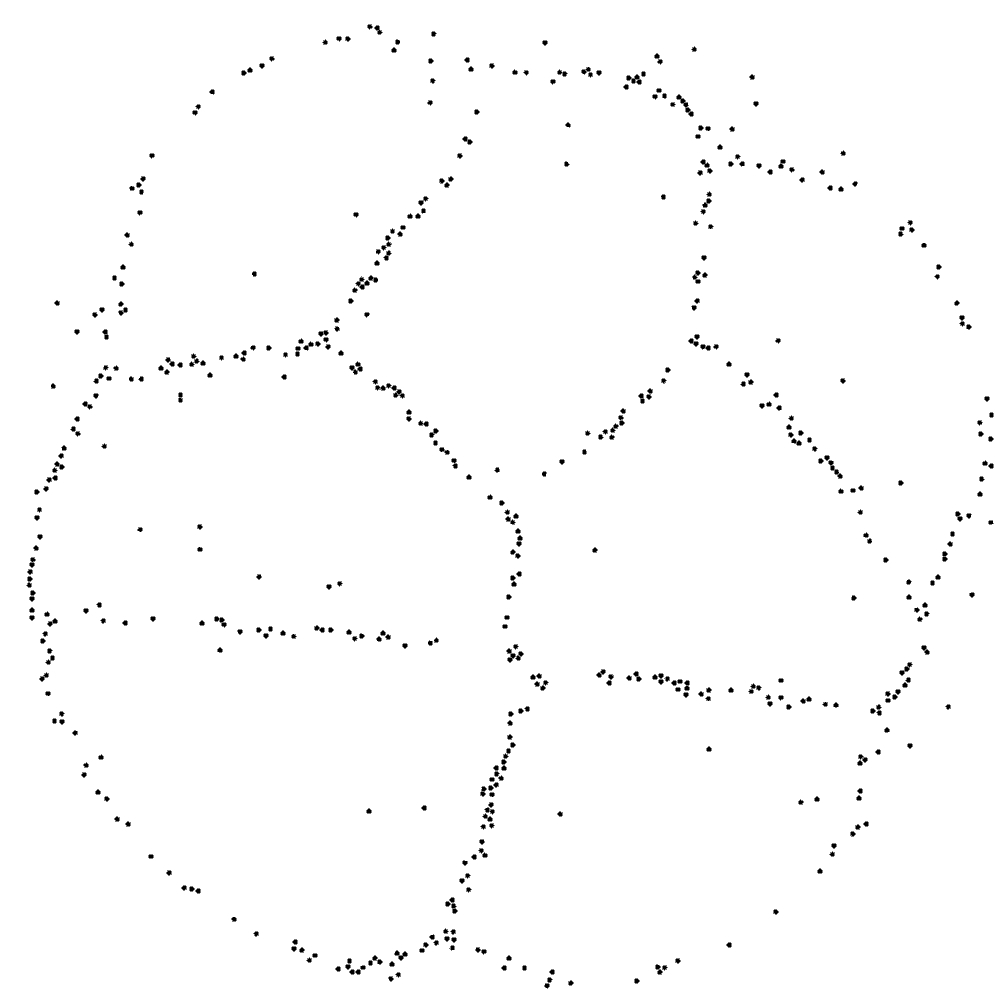}}
\subfigure[$\eta=0.4$]{\includegraphics[width=0.24\textwidth]{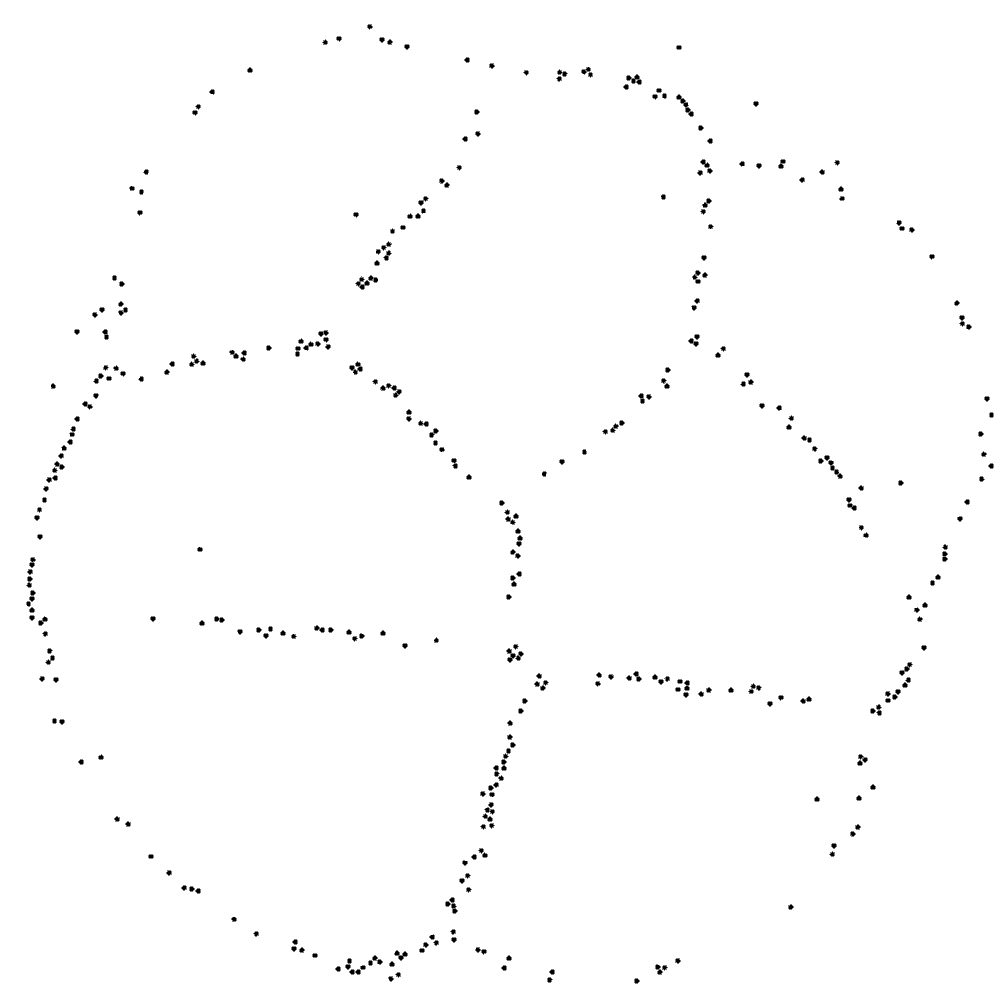}}
\subfigure[$\eta=0.425$]{\includegraphics[width=0.24\textwidth]{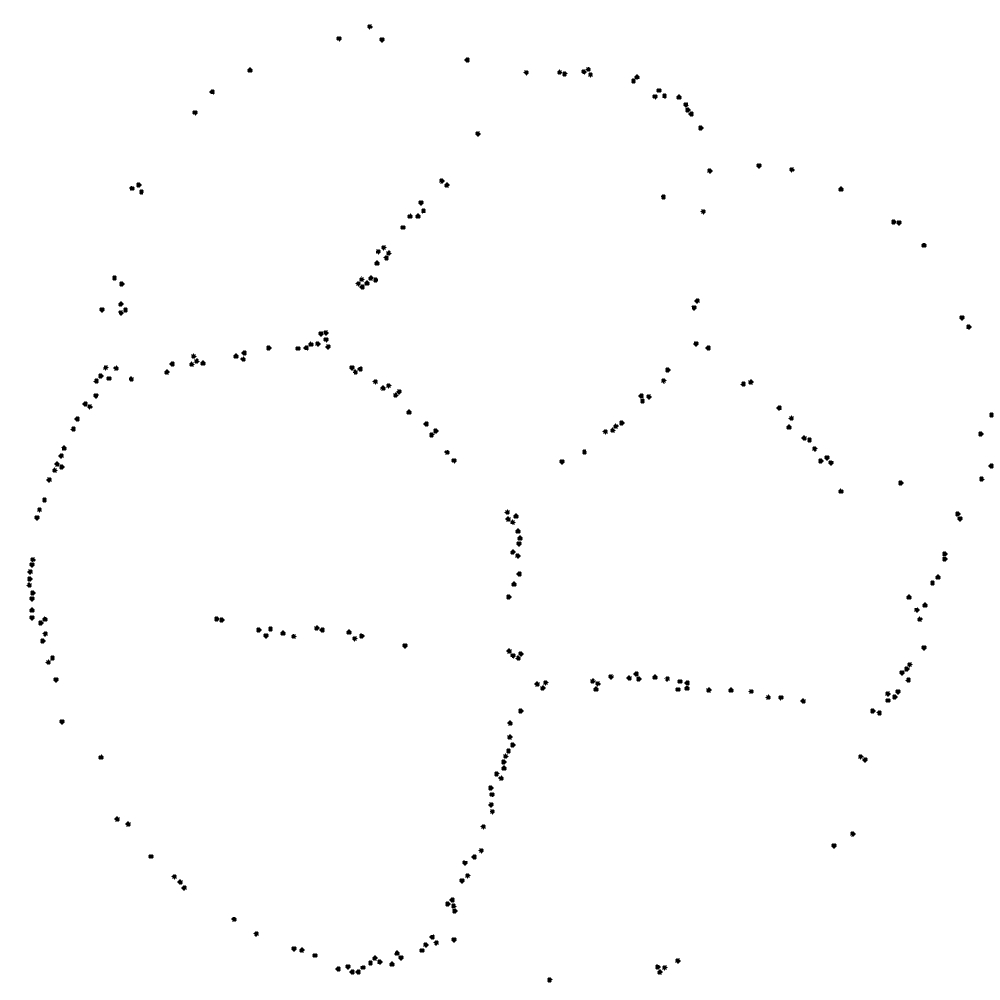}}
\subfigure[$\eta=0.45$]{\includegraphics[width=0.24\textwidth]{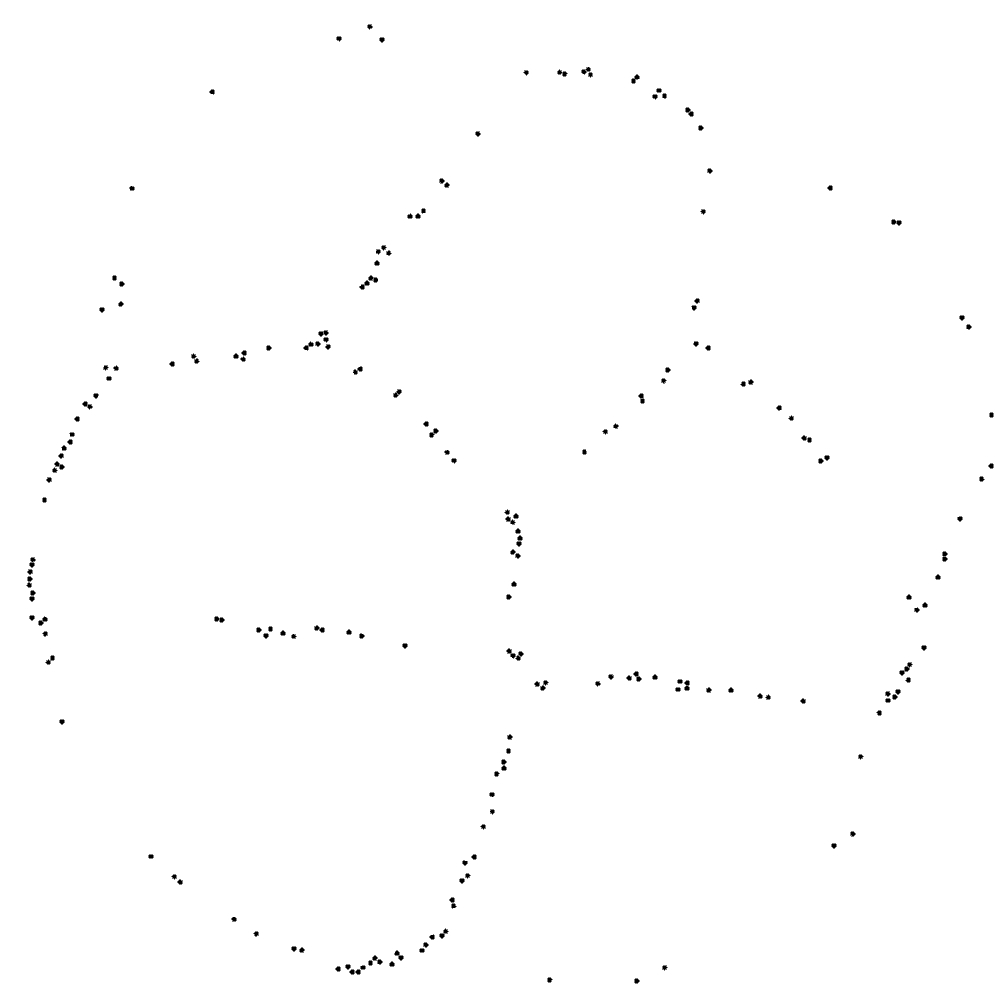}}
\subfigure[$\eta=0.4755$]{\includegraphics[width=0.24\textwidth]{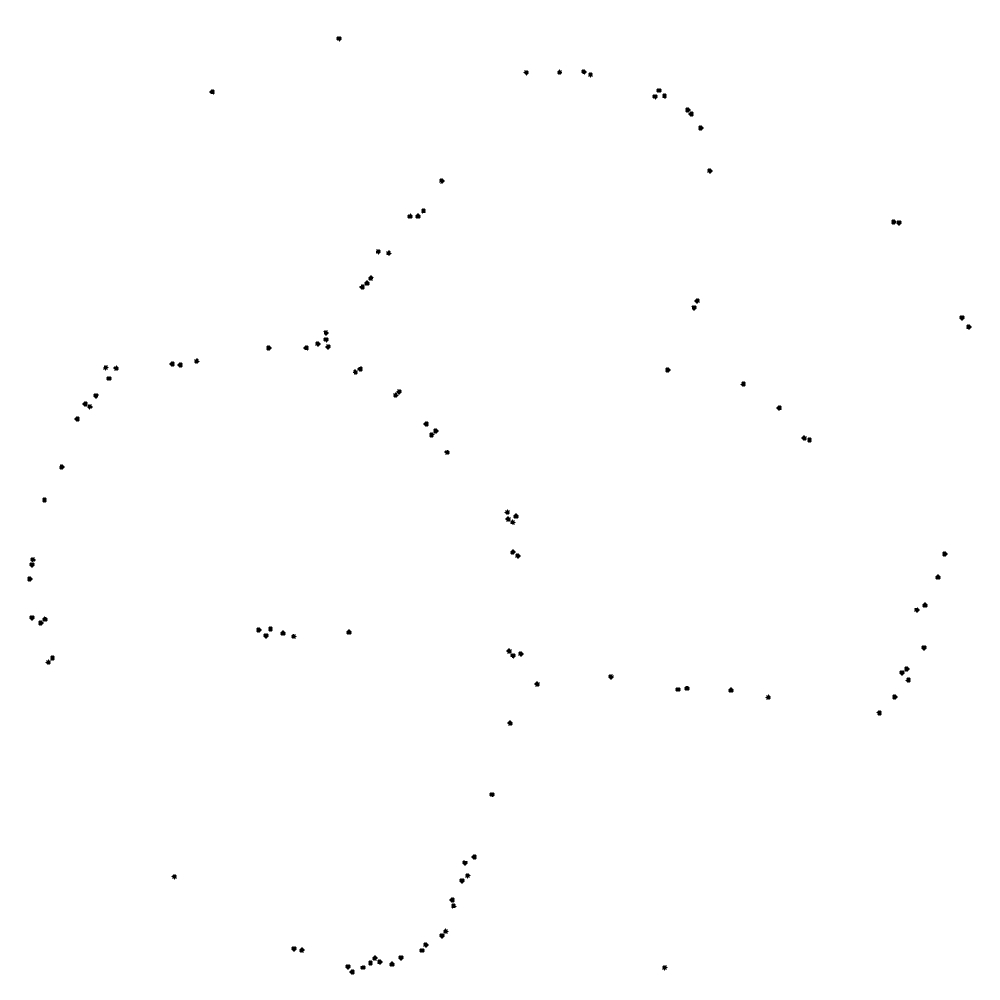}}
\subfigure[$\eta=0.5$]{\includegraphics[width=0.24\textwidth]{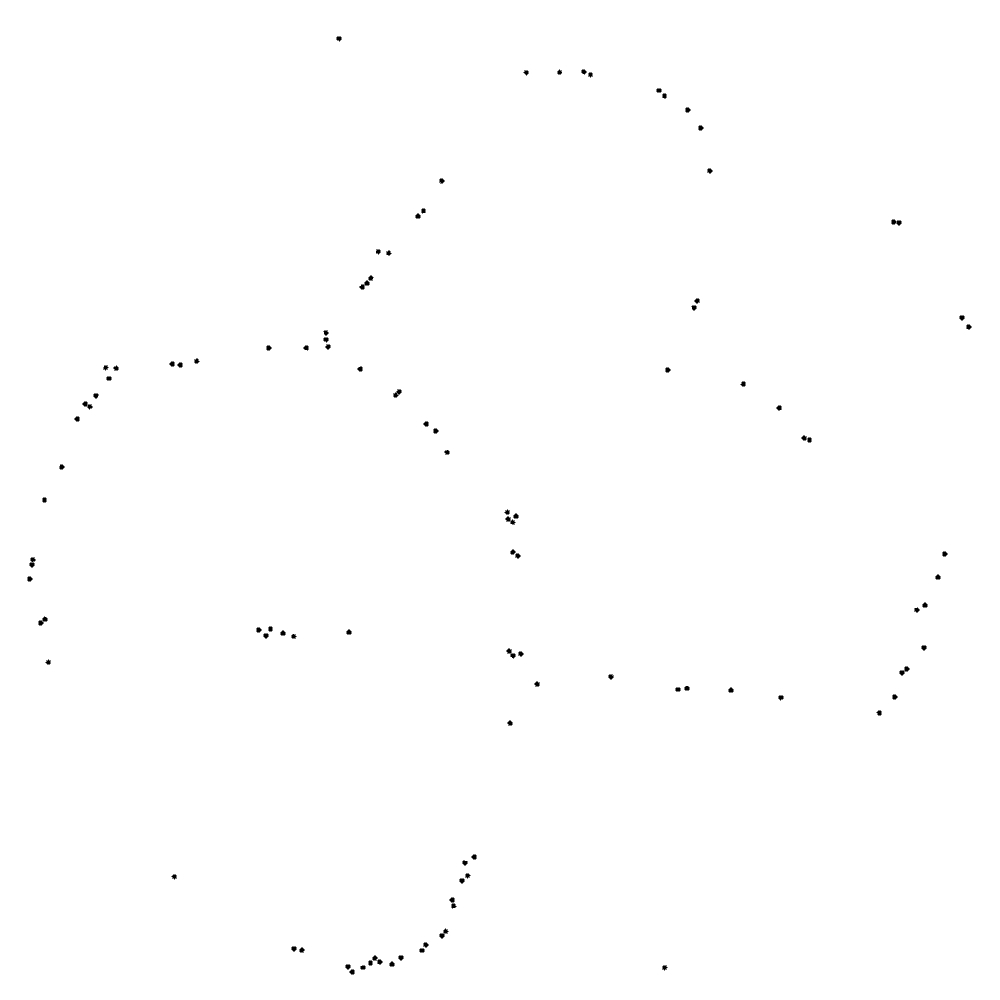}}
\subfigure[$\eta=0.525$]{\includegraphics[width=0.24\textwidth]{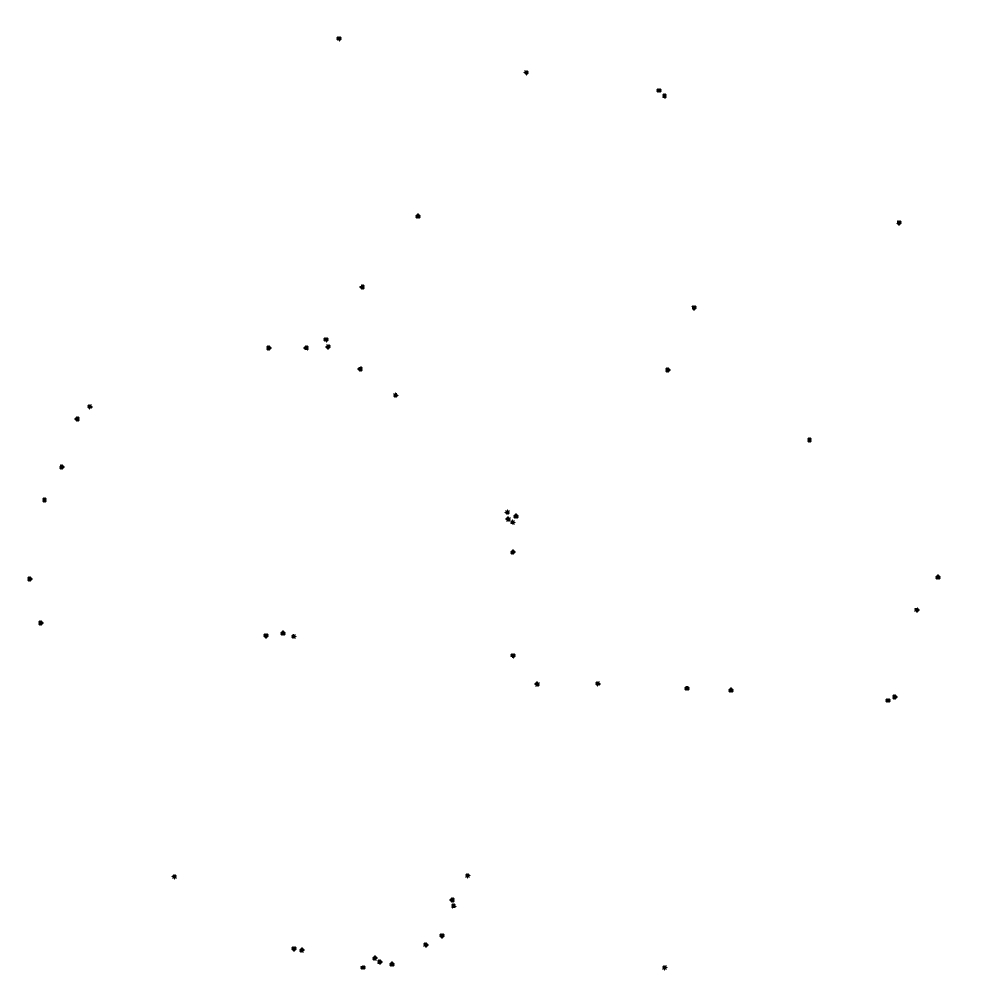}}
\caption{Quality of approximation depends on $\eta$. (a)~Voronoi diagram  $\mathfrak{V}(\mathbf{B})$ of 
seven planar points set $\mathbf B$ constructed by Fortuna's sweepline algorithm, (b)--(o)~stationary configurations of
precipitate in automaton $\mathcal V$ excited by $\mathbf B$ for different $\eta= 0.2 \ldots 0.525$. In snapshots (b)--(o)
black pixels symbolise Voronoi cells of $\mathcal V$ in precipitate state, cells in refractory states are blank.}
\label{parameterisation}
\end{figure}

Even in a single example shown in Fig.~\ref{points} it is clear that $\mathcal V$ does not 
calculate $\mathfrak{V}(\mathbf{B})$ precisely, it rather approximates it with discrete domains of precipitation, and 
also introduces some noise (sites of precipitation not coinciding with bisectors of $\mathfrak{V}(\mathbf{V})$).  
Only threshold of precipitation $\eta$ can be varied in our model. How are a degree of approximation and noise to 
signal ratio depends on value of $\eta$? To answer we undertook a series of computational experiments, illustrated -- for some values of $\eta$ --- in Fig.~\ref{parameterisation}.

\begin{figure}[!tbp]
\centering
\subfigure[]{\includegraphics[width=0.49\textwidth]{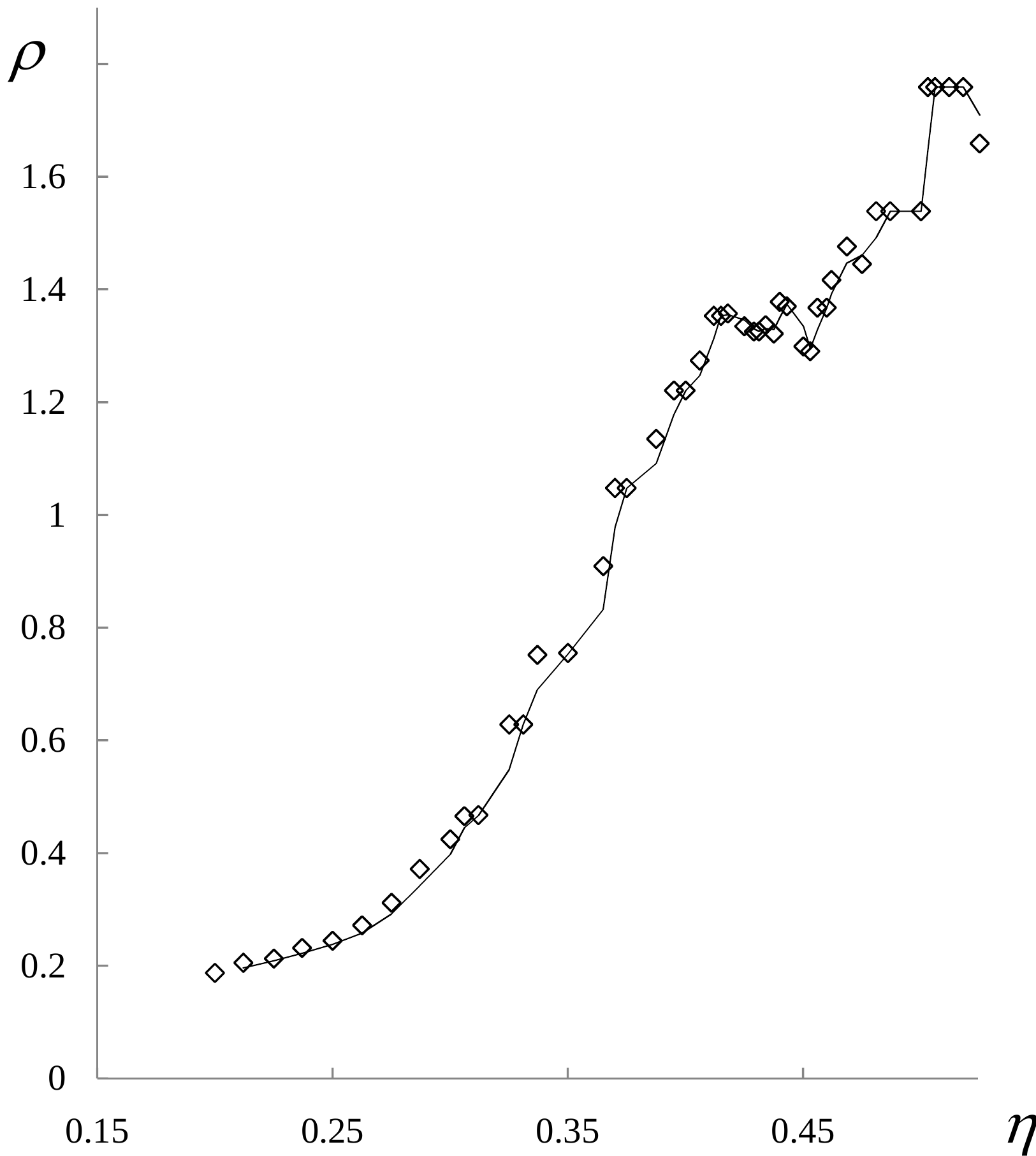}}
\subfigure[]{\includegraphics[width=0.49\textwidth]{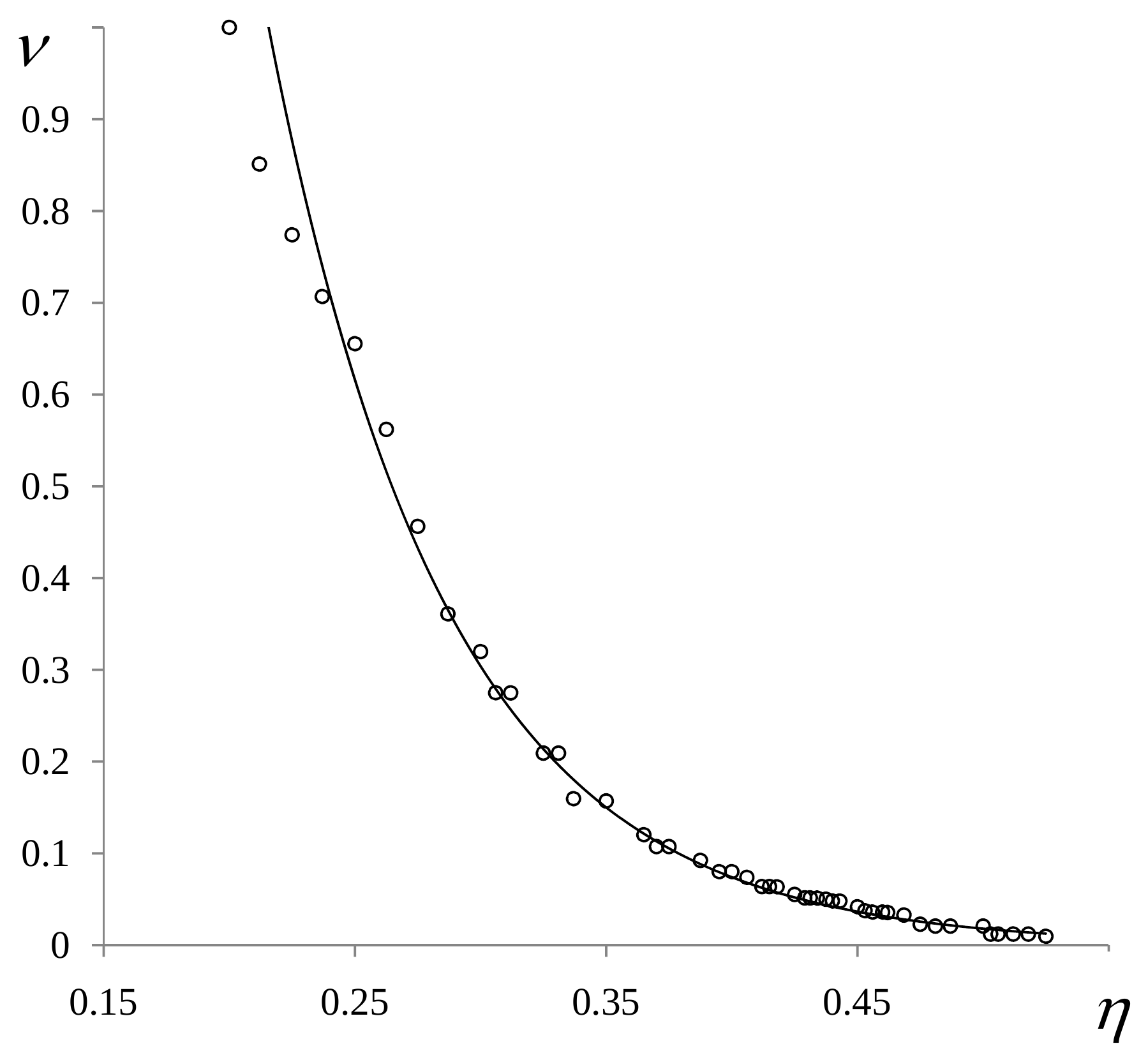}}
\caption{Characterisation of approximation of $\mathfrak{V}(\mathbf{B})$ (Fig.~\ref{parameterisation}a) 
by automaton $\mathcal V$: (a)~$\rho$ vs $\eta$ and (b)~$\nu$ vs $\eta$. Markers (rhombs and circles) show
actual values of $\nu(\eta)$ and $\rho(\eta)$, while solid lines are moving average trend line purely for eye guidance.}
\label{graphs}
\end{figure}

For a set $\mathbf B$ of seven points we constructed 'classical' planar Voronoi diagram $\mathfrak{V}(\mathbf{B})$
(Fig.~\ref{parameterisation}a). Then for $\eta= 0.2 \ldots 0.525$ we excited $\mathcal V$ with $\mathbf B$, waited till $\mathcal V$ reaches its stationary configuration, where all cells are in either precipitate or refractory state, and compared the configurations of $\mathcal V$ and image of $\mathfrak{V}(\mathbf{B})$. For each $\eta$ we recorded a fraction $\rho$ of precipitate-cells in a final configuration of $\mathcal{V}(\eta)$ to a total number of precipitate-cells in the same 
configuration (Fig.~\ref{graphs}a).

We found that while density $\nu$, calculated as a ratio of a number of precipitate-cells in stationary configuration $\mathcal{V}(\eta)$ to a number of precipitate-cells in configuration of $\mathcal{V}(0.2)$, 
exponentially decreases with increase of precipitation threshold $\eta$ (Fig.~\ref{graphs}b) degree of approximation $\rho$ 
shows more intriguing behaviour (Fig.~\ref{graphs}a). From $\eta=0.2$ till $\eta=0.41$ `signal-to-noise' ratio $\rho(\eta)$ grows 
almost exponentially.  Trend of $\rho(\eta)$ undergoes S-type transition between $\eta=0.41$ and $\eta=0.453$ with one pick $\rho(0.44)$ and two cavities $\rho(0.43)$ and $\rho(0.452)$. As illustrated in Fig.~\ref{parameterisation} $\mathcal V$ with 
precipitation threshold in the zone around $\eta=0.4$, onset  of `strange' behaviour of $\rho(\eta)$), 
demonstrates best approximation of $\mathfrak{V}(\mathbf{B})$, as recognised by unaided eye.

\section{Arbitrary-shaped planar objects and contours}
\label{shapes}

\begin{figure}[!tbp]
\subfigure[data]{\includegraphics[width=0.49\textwidth]{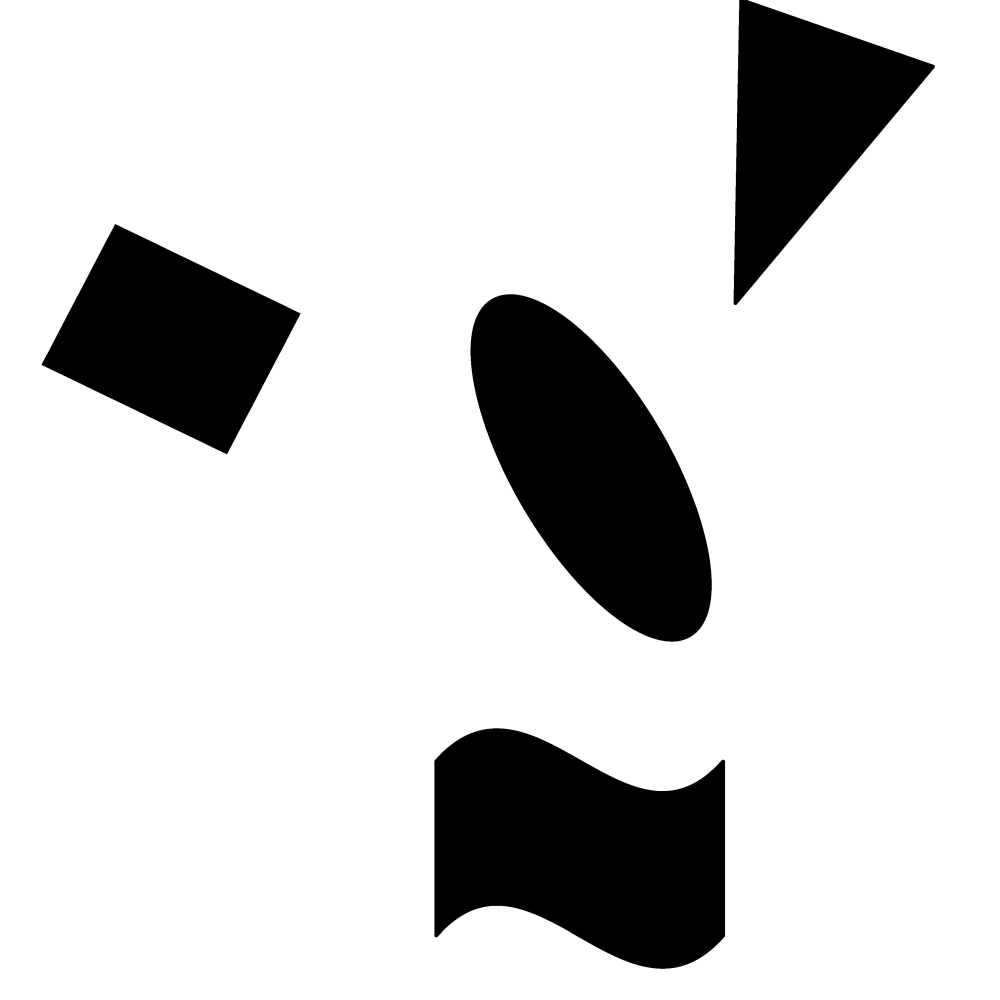}}
\subfigure[$t=1$]{\includegraphics[width=0.49\textwidth]{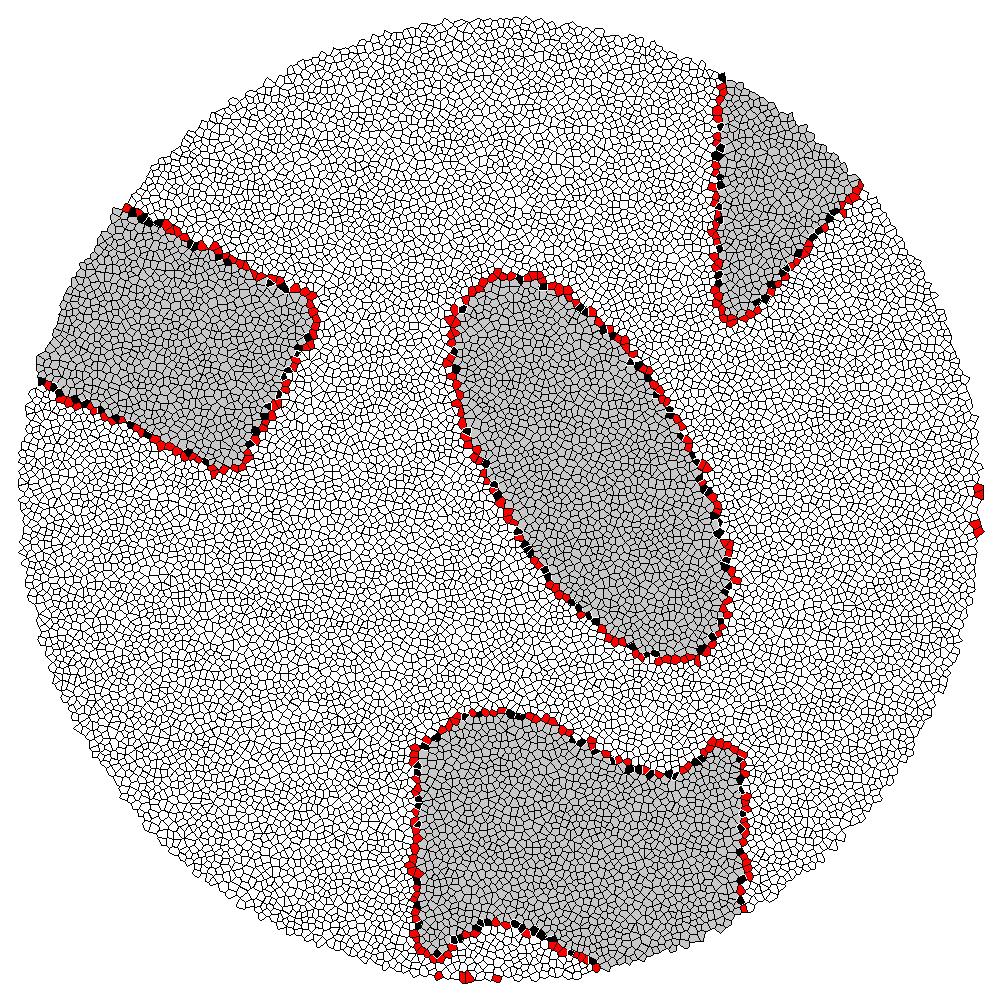}}
\subfigure[$t=5$]{\includegraphics[width=0.49\textwidth]{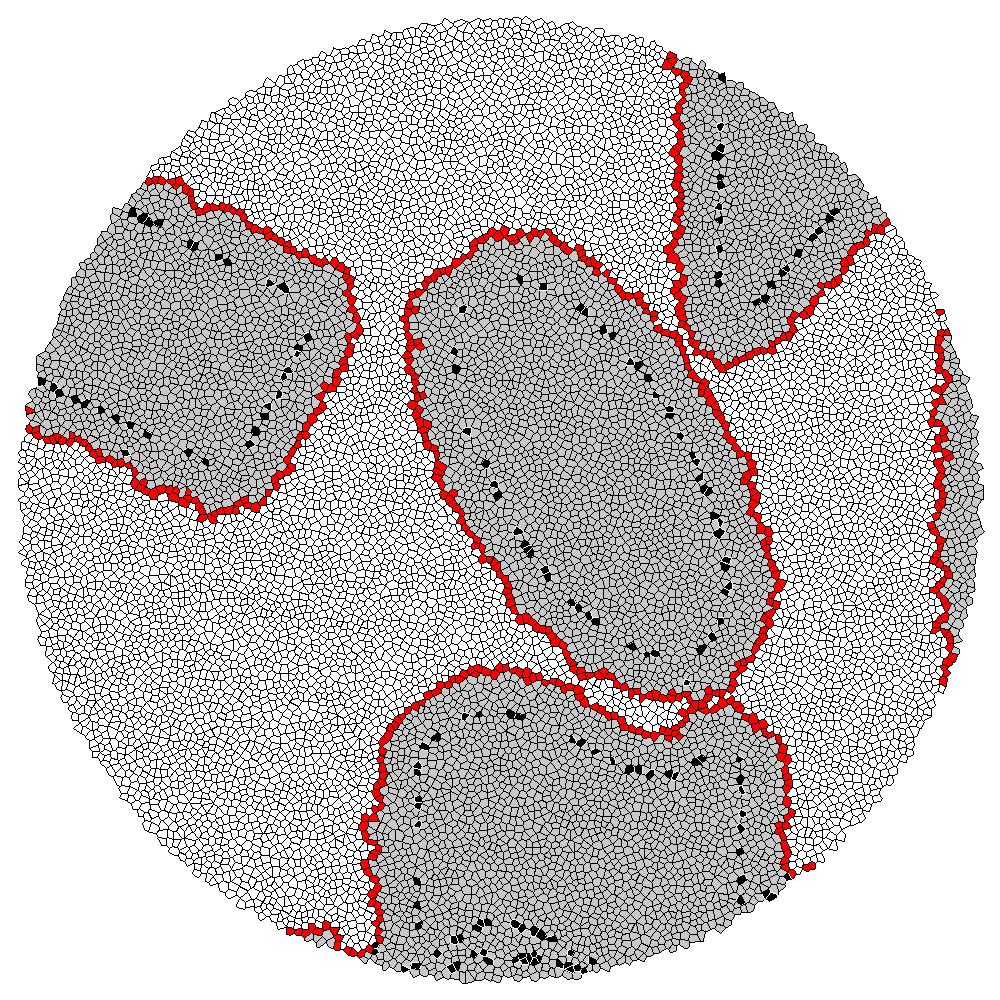}}
\subfigure[$t=12$]{\includegraphics[width=0.49\textwidth]{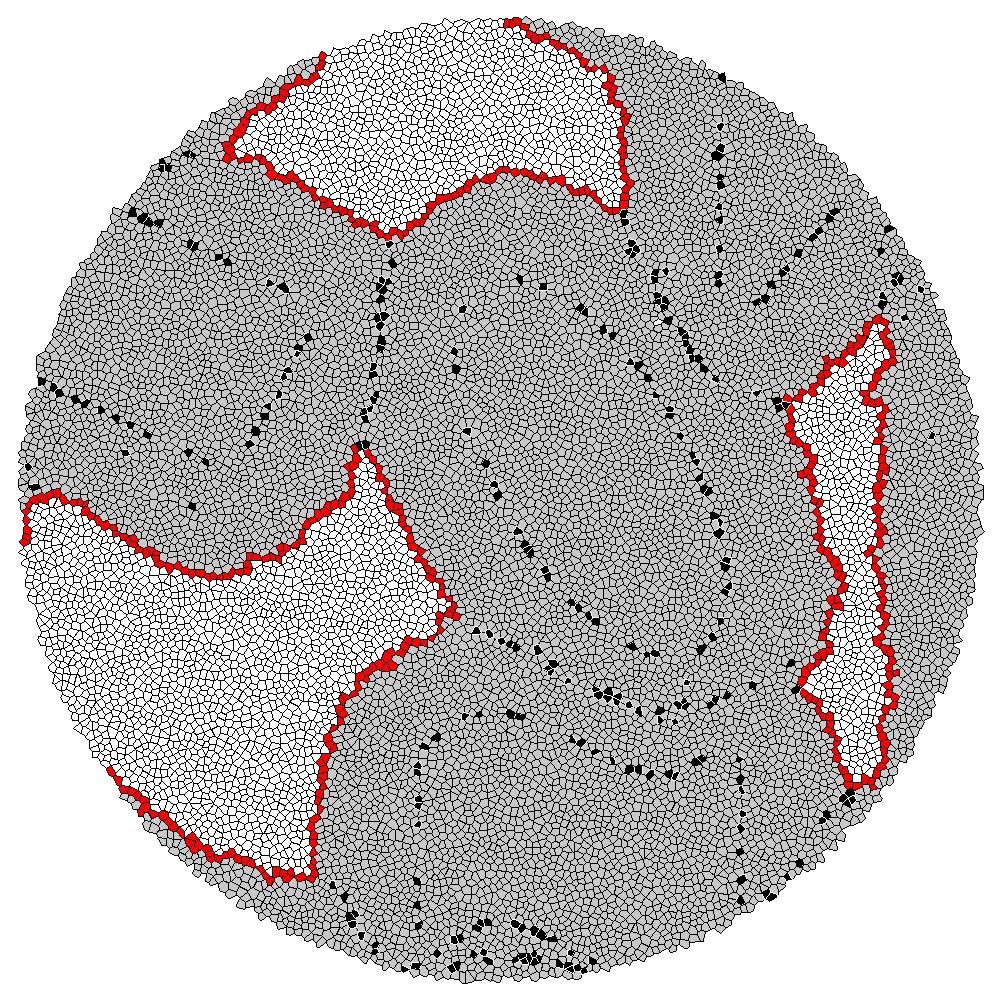}}
\subfigure[$t=20$]{\includegraphics[width=0.49\textwidth]{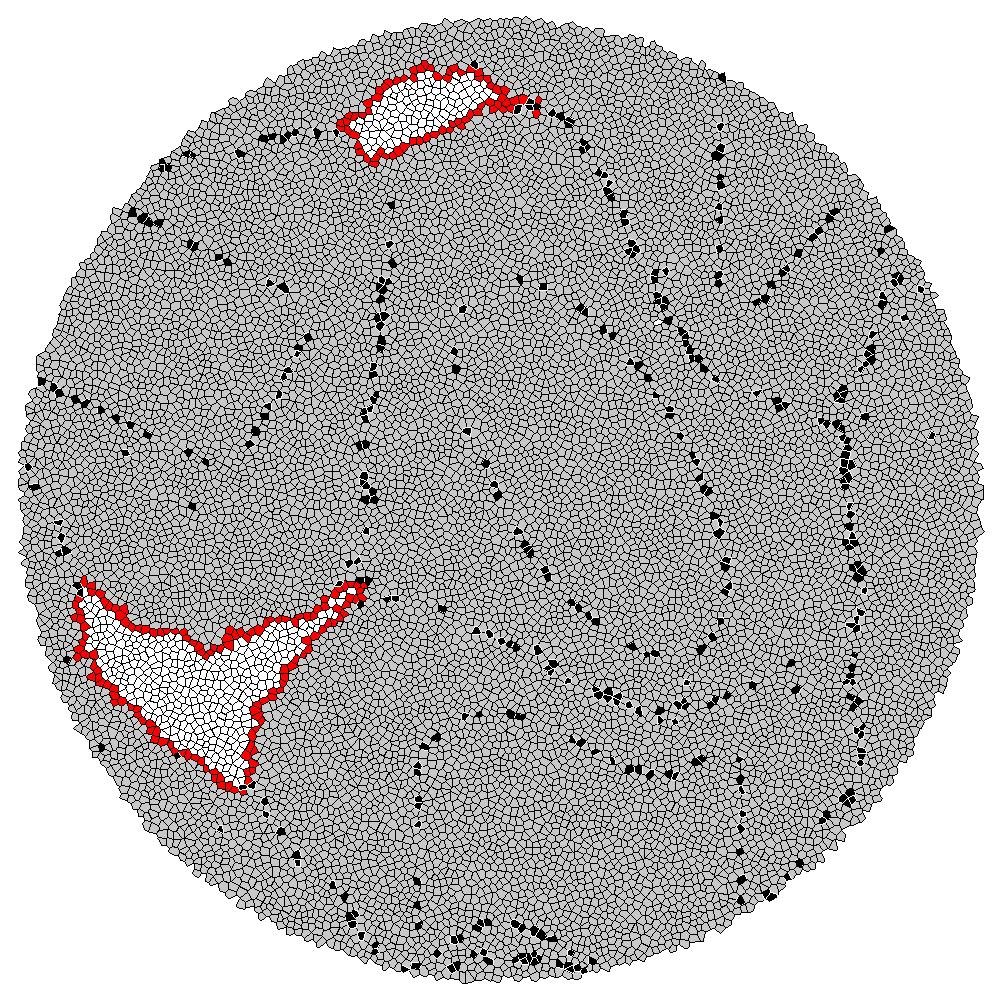}}
\subfigure[$t=28$]{\includegraphics[width=0.49\textwidth]{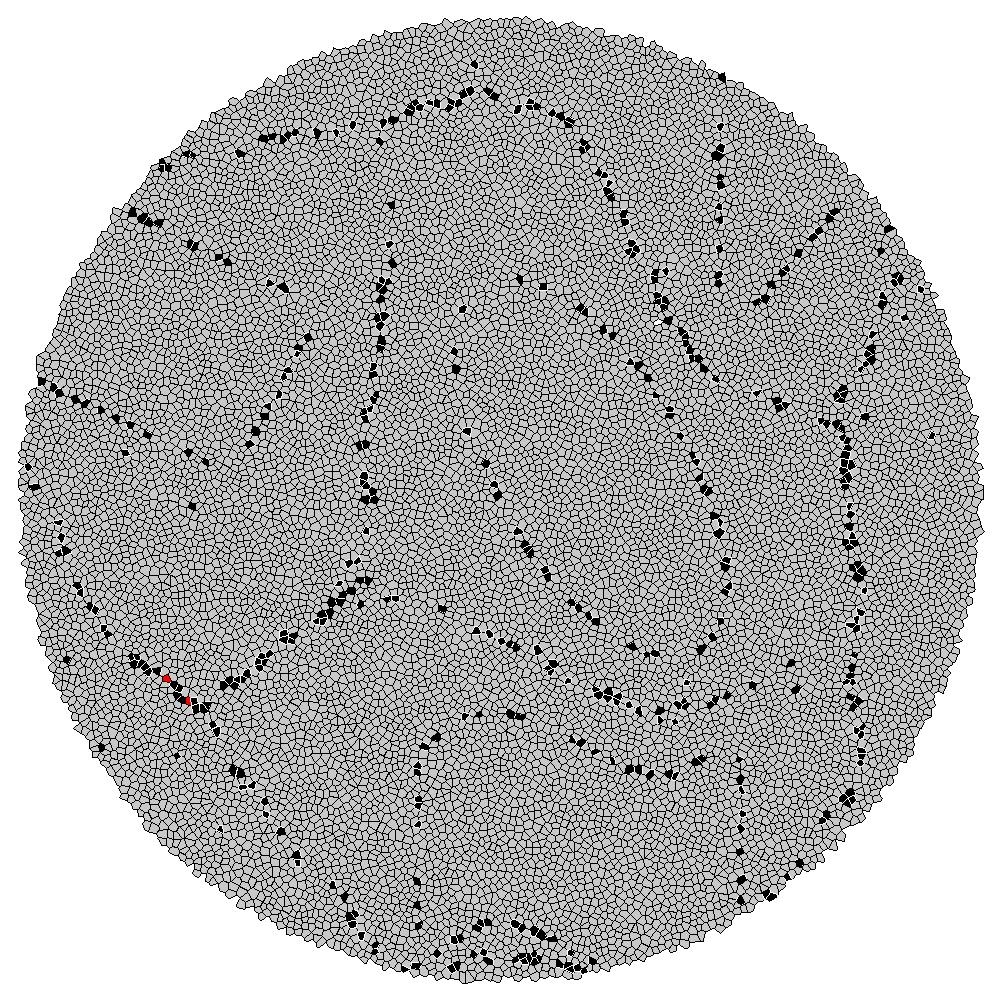}}
\caption{Approximation of Voronoi diagram $\mathfrak{V}(\mathbf{O})$ of planar shapes $\mathbf O$ in 
Voronoi automaton $\mathcal V$. Voronoi cell of $\mathcal V$ in precipitate state are black, cells in refractory states are blank. Excited cells are red (gray). (a)~representation of $\mathbf O$,
(b)--(f)~development of $\mathcal V$. $\mathfrak{V}(\mathbf{O})$ is represented by black pixels in (f).} 
\label{objects}
\end{figure}

\begin{figure}[!tbp]
\subfigure[data]{\includegraphics[width=0.32\textwidth]{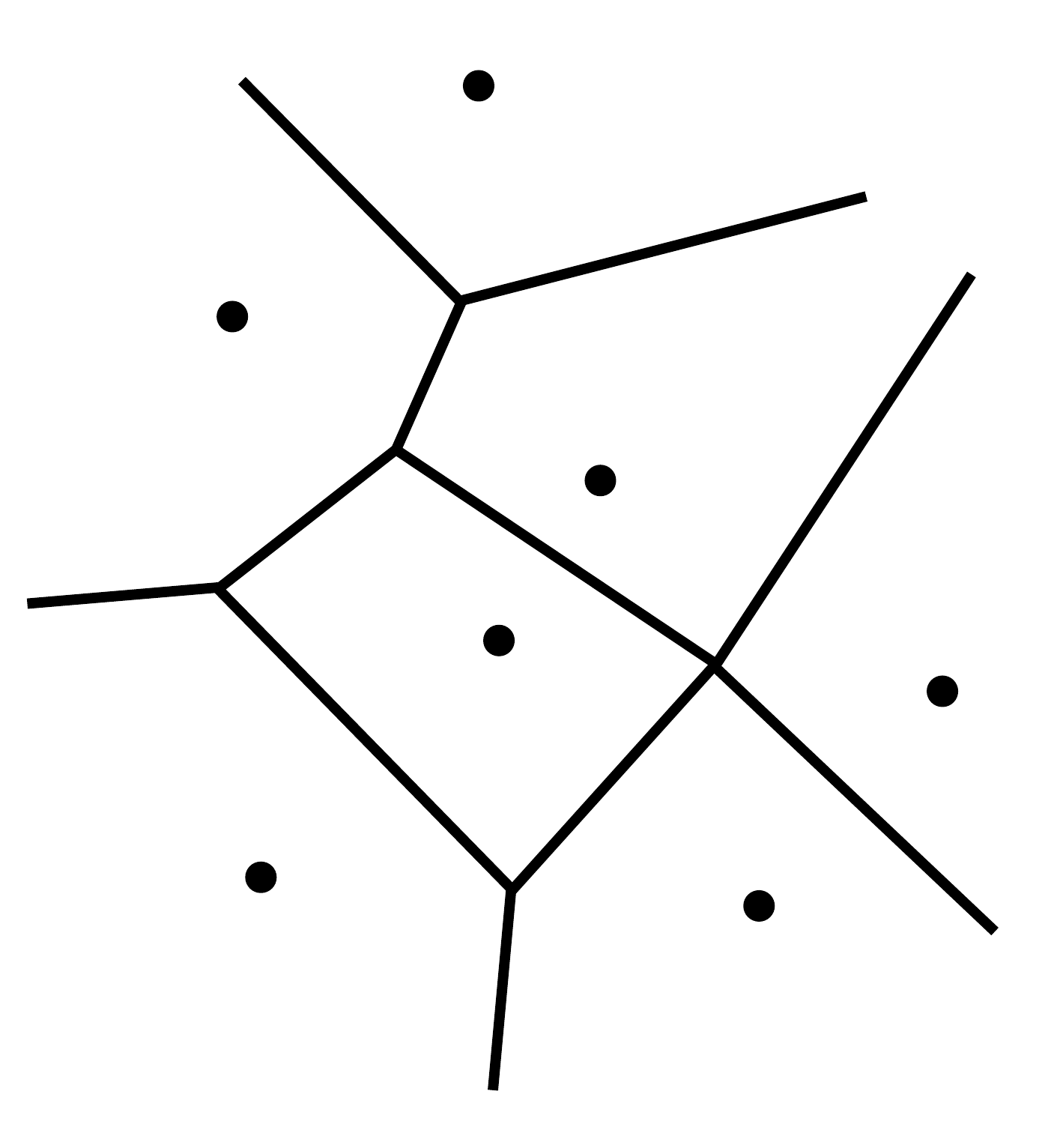}}
\subfigure[$t=1$]{\includegraphics[width=0.32\textwidth]{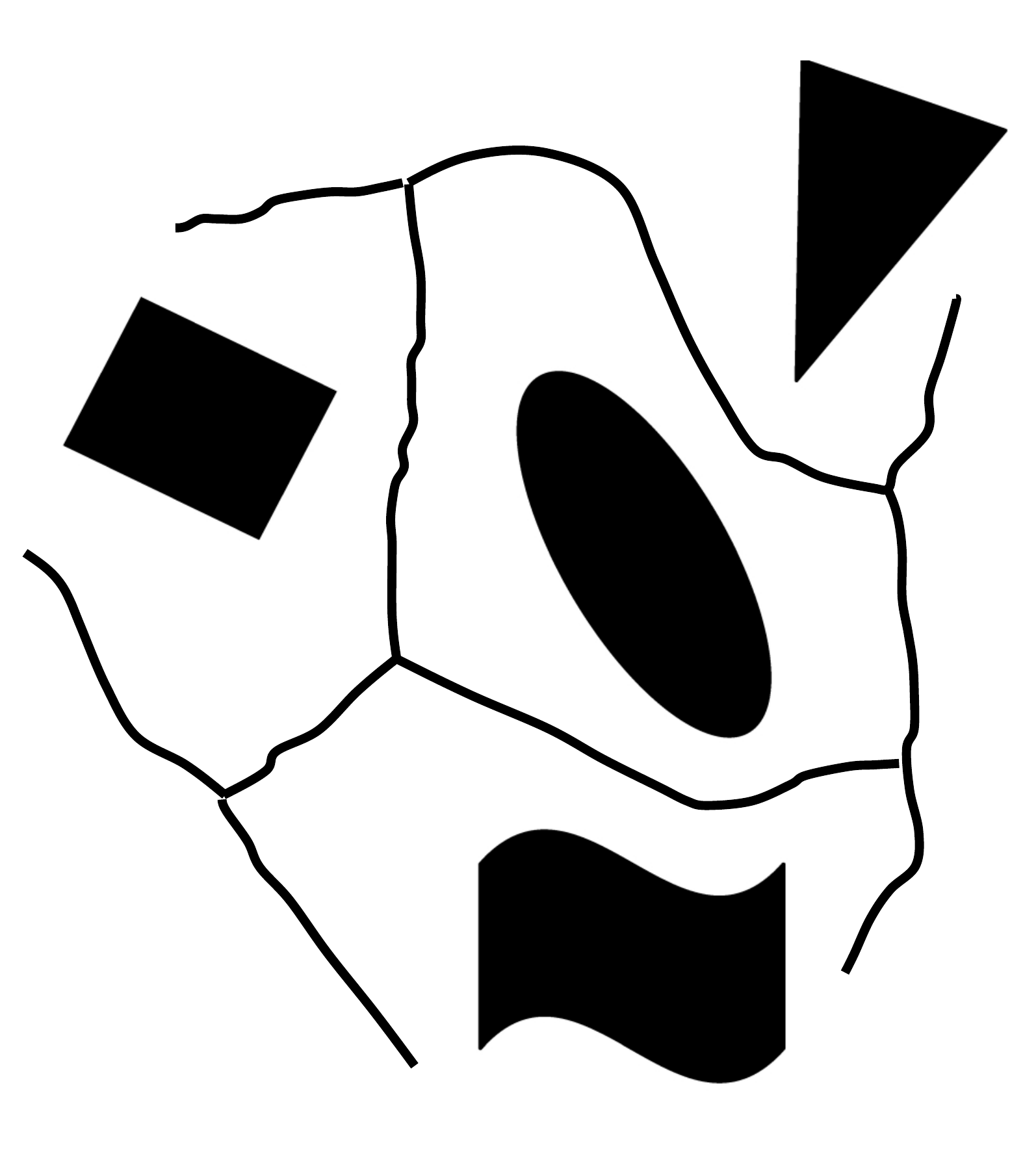}}
\subfigure[$t=5$]{\includegraphics[width=0.32\textwidth]{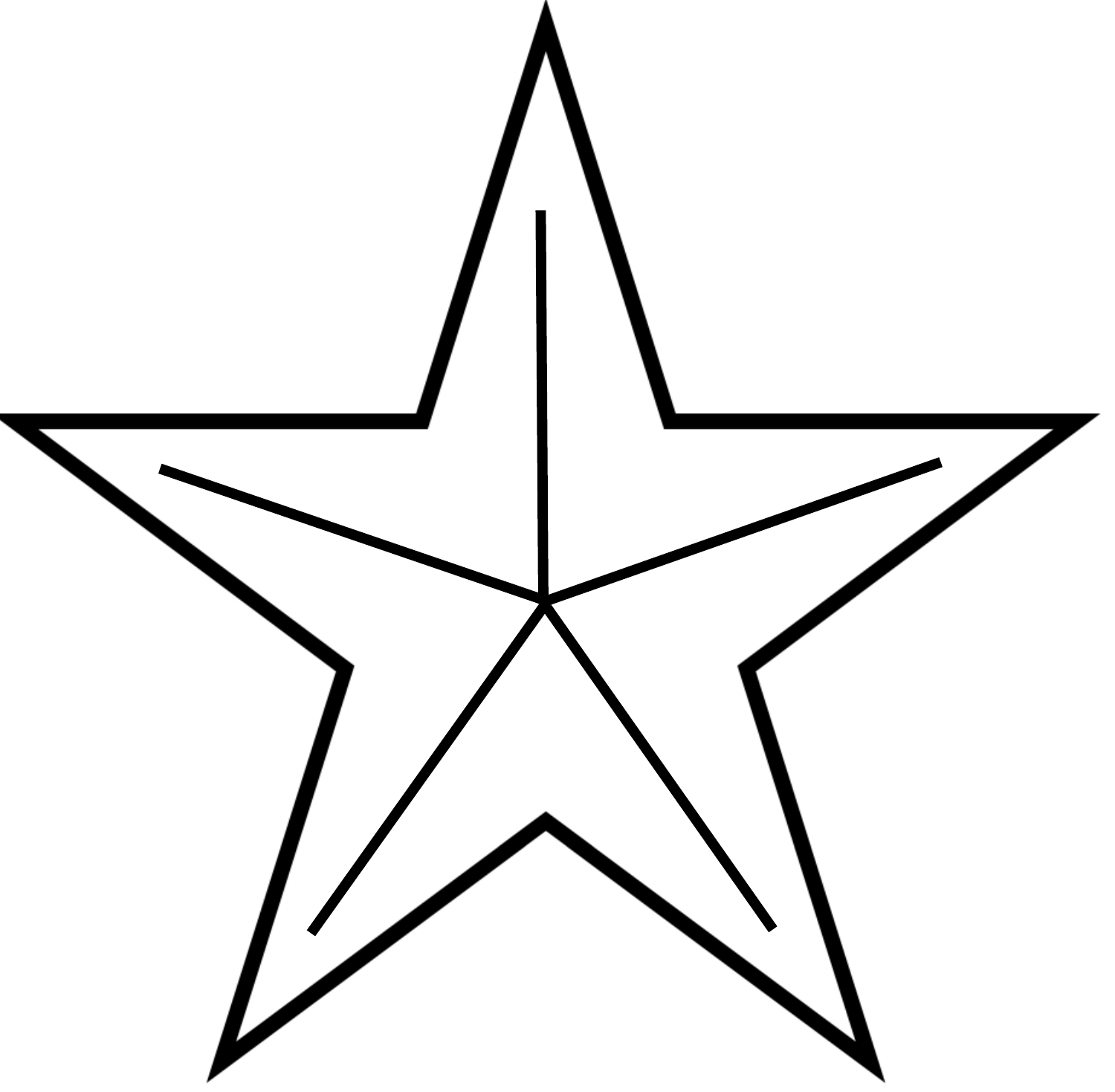}}
\caption{Schematic representation of data objects and their Voronoi diagram or skeleton
for precipitation-based approximation shown in (a)~Fig.~\ref{points}, 
(b)~Fig.~\ref{objects}, and (c)~Fig.~\ref{skeleton}.} 
\label{scheme}
\end{figure}

The Voronoi automaton $\mathcal V$ copes well with data set $\mathbf O$ of planar finite shapes. To approximate 
diagram $\mathfrak V$ of arbitrary planar shapes we project the objects of $\mathbf O$ (Fig.~\ref{objects}a)
onto $\mathcal V$ in such a manner that at time $t=0$ all cells of $\mathcal V$ covered (coinciding with pixels of)
by shapes from $\mathbf O$ becomes excited (Fig.~\ref{objects}b).  Waves of excitation travel across $\mathcal V$
and provoke precipitation during their collisions (Fig.~\ref{objects}c--e).  When all cells take refractory or 
precipitate state the automaton's configuration becomes stationary. The spatial distribution of precipitate 
represents Voronoi diagram of data shapes (see Fig.~\ref{objects}f and Fig.~\ref{scheme}b). A precipitate-enhancement of 
original position of date shape is a byproduct of $\mathcal V$'s development.

\begin{figure}[!tbp]
\subfigure[data]{\includegraphics[width=0.49\textwidth]{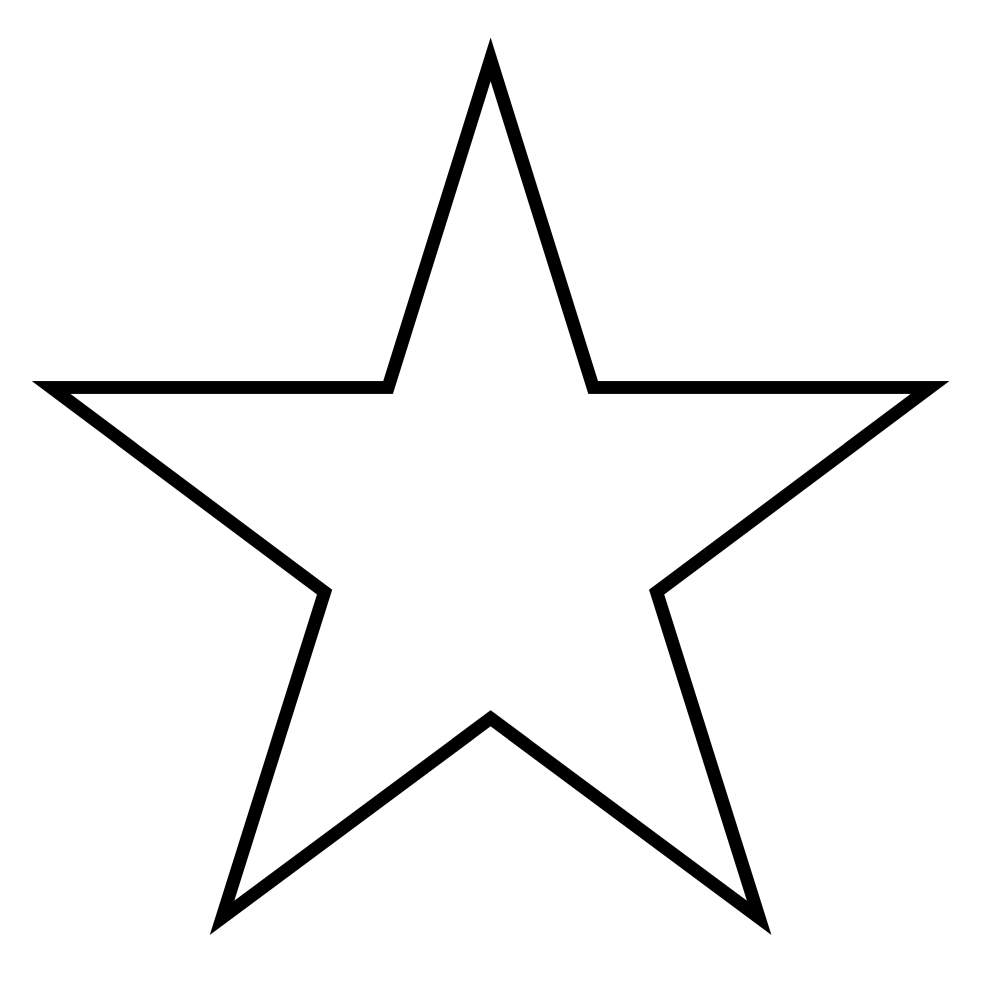}}
\subfigure[$t=1$]{\includegraphics[width=0.49\textwidth]{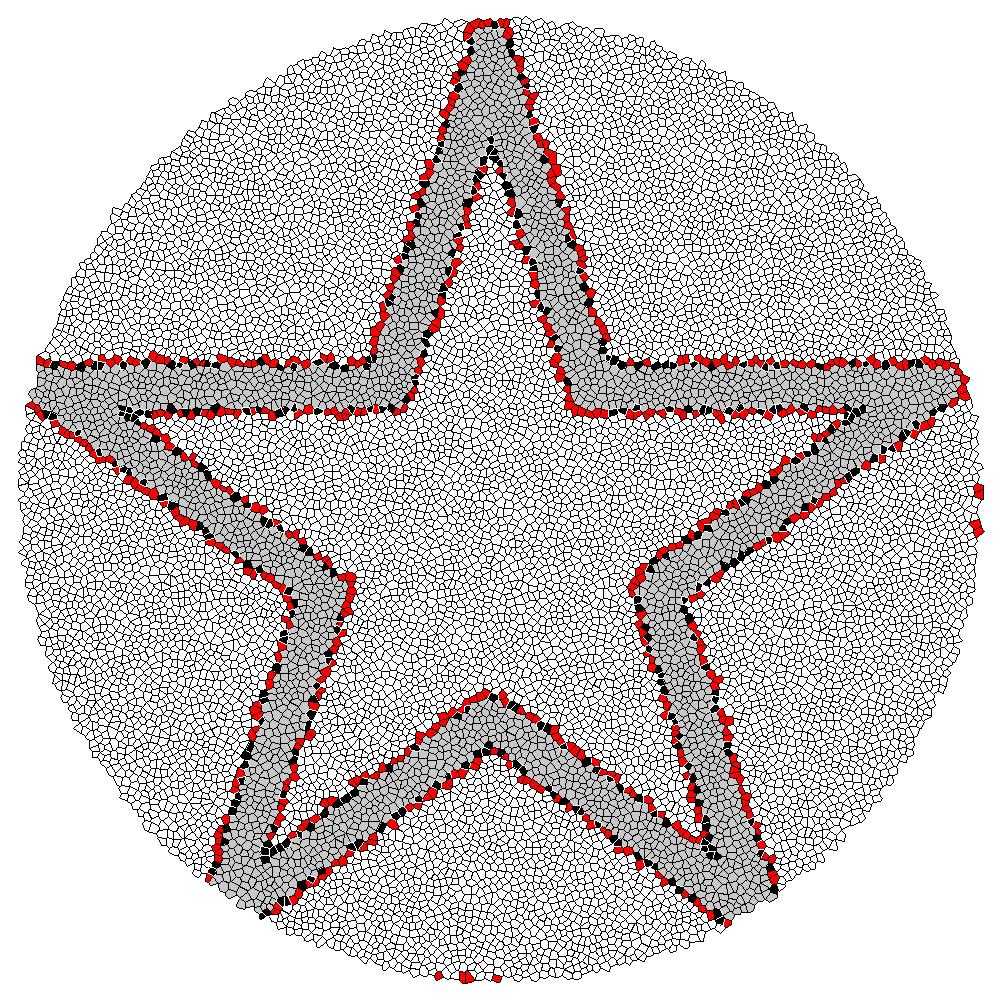}}
\subfigure[$t=5$]{\includegraphics[width=0.49\textwidth]{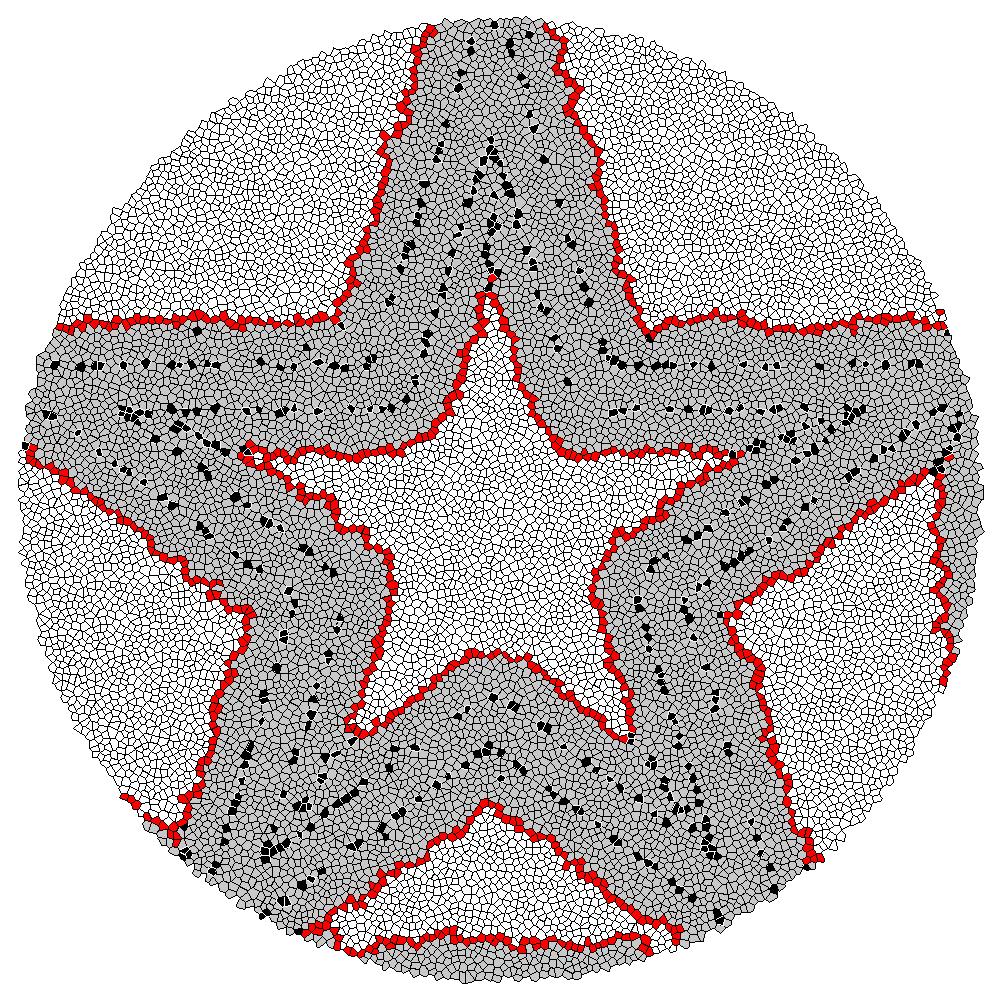}}
\subfigure[$t=13$]{\includegraphics[width=0.49\textwidth]{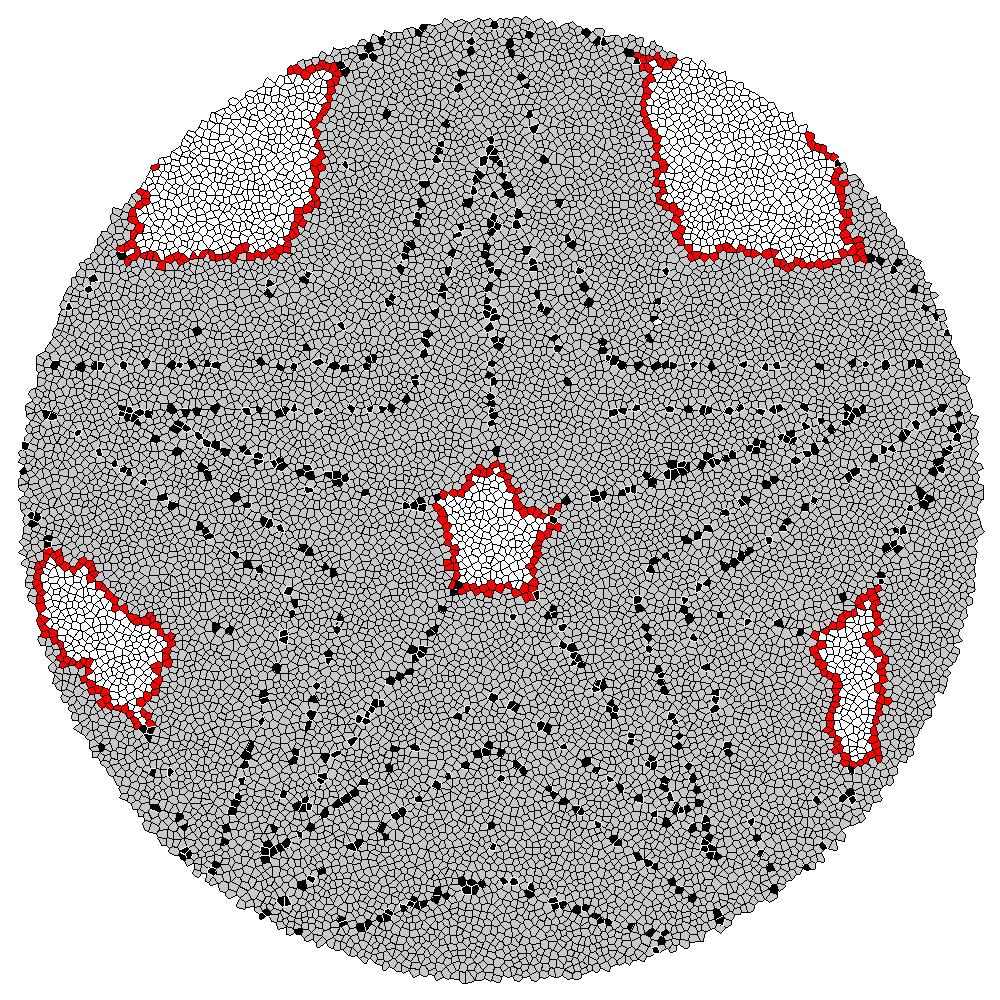}}
\subfigure[$t=26$]{\includegraphics[width=0.49\textwidth]{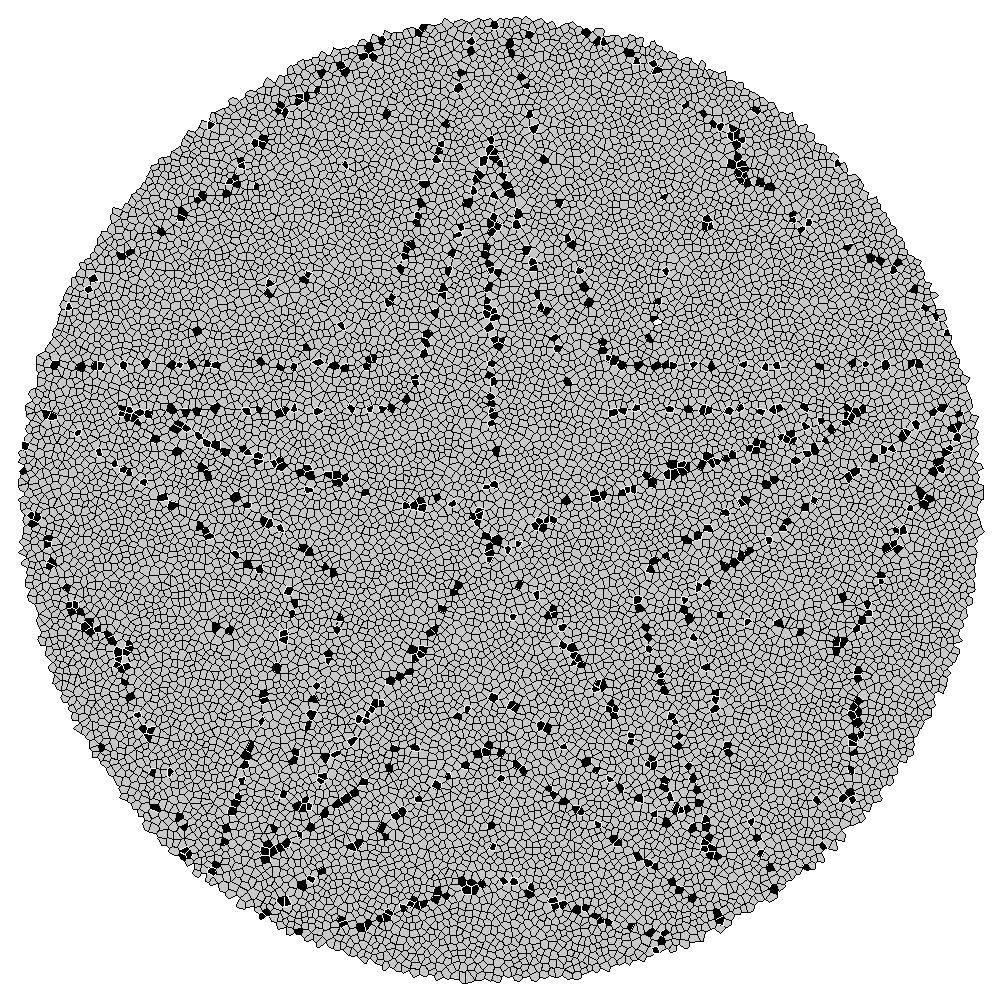}}
\caption{Approximation of a skeleton of a star-contour $\mathbf S$ by $\mathcal V$.  Voronoi cell of $\mathcal V$ in precipitate state are black, cells in refractory states are blank. Excited cells are red (gray). (a)~representation of a star-contour,
(b)--(e)~development of $\mathcal V$.} 
\label{skeleton}
\end{figure}

A skeleton of a planar contour is a set of centers of bitangent circles lying inside the
contour~\cite{calabi_1968}.  Blum's grass-fire algorithm for computing skeleton employs 
propagating patterns~\cite{blum_1967,blum_1973,calabi_1968}: to compute a skeleton we 
set a contour on fire and let the fire spread, wuench points where the advancing fire-fronts 
collide represent the skeleton. Most known algorithms of skeletonisation are based on 
Blum's approach: simulation of grass-fire~\cite{leymarie_1992}, distance transform~\cite{rosenfeld_1968},
analytical construction of medial axis and topological thinning~\cite{pearce_1993,attali_1997}.

To approximate a skeleton of a shape $\mathbf S$ (Fig.~\ref{skeleton}a) we project $\mathbf S$ onto $\mathcal V$. Voronoi cells of $\mathcal V$ corresponding to black pixels of $\mathbf S$ are excited (Fig.~\ref{skeleton}b). Excitation waves propagate inside and outside of the domain $\mathbf S$ of initial excitation; we can ignore behaviour of outward waves. Waves travelling inside the contour $\mathbf S$ trigger precipitation at the sites of the waves' interaction (Fig.~\ref{skeleton}c--e). Distribution of 
precipitate (Fig.~\ref{skeleton}e) inside $\mathbf S$ approximates skeleton of $\mathbf S$ (Fig.~\ref{scheme}c).

\newpage

\section{Summary}
\label{discussion}

We demonstrated that is possible to solve some problems of computational geometry on a discrete model of irregularly arrangements non-uniform vesicles filled with  chemical mixtures~\cite{adamatzky_BZBALLS_2010, adamatzky_polymorphic_2010, holley_2010}. We shown that a threshold of relative local excitation density can be used to  parameterize excitation and precipitation dynamics, enhance results of the computation and reduce noise-to-signal ratio. We proved that the same optimal threshold of precipitation works well not only for approximating Voronoi diagram of planar sets but also for 
arbitrary shapes and skeletonisation of planar contours. In terms of automata-network based computation we advanced our previous results on excitation in Delaunay automata~\cite{adamatzky_delaunay} and $\beta$-skeletons~\cite{adamatzky_betaskeletons,alonsosanz_betaskeletons}. The parameterisation developed could be used in the 
designing experimental laboratory prototypes of fine-grained compartmentalised excitable chemical processors, e.g. using micro-emulsion approach~\cite{kaminaga_2006}, and developing nano-scale massively parallel computers~\cite{bandyo_2010}.

\section{Acknowledgment}

The work is part of the European project 248992 funded under 7th FWP (Seventh Framework Programme) FET Proactive 3: Bio-Chemistry-Based Information Technology CHEM-IT (ICT-2009.8.3).

\end{document}